\documentclass[prl,aps,twocolumn,amsmath,amssymb,floatfix,pra,reprint,footinbib,superscriptaddress,showpacs,longbibliography]{revtex4-1}

\usepackage{graphics}      
\usepackage{graphicx}      
\usepackage{longtable}     
\usepackage{url}           
\usepackage{bm}    
\usepackage{braket}        
\usepackage{xcolor}
\usepackage{float}  
\usepackage{natbib}
\usepackage{graphicx}
\usepackage{comment}
\usepackage{siunitx}
\usepackage{appendix}
\usepackage{amsmath,amssymb,epsfig,color}

\newcommand{\bea}{\begin{eqnarray}}
\newcommand{\eea}{\end{eqnarray}}

\usepackage{changes}

\begin{document}

\title{Many-Body Dephasing in a Trapped-Ion Quantum Simulator}

\author{Harvey B. Kaplan}
\thanks{H.K. and L.G. contributed equally to this work.}
\affiliation{Joint Quantum Institute, Department of Physics and Joint Center for Quantum Information and Computer Science, University of Maryland, College Park, MD 20742, USA}

\author{Lingzhen Guo}
\thanks{H.K. and L.G. contributed equally to this work.}
\affiliation{Max Planck Institute for the Science of Light, Staudtstrasse 2, 91058 Erlangen, Germany}

\author{Wen Lin Tan}
\affiliation{Joint Quantum Institute, Department of Physics and Joint Center for Quantum Information and Computer Science, University of Maryland, College Park, MD 20742, USA}

\author{Arinjoy De}
\affiliation{Joint Quantum Institute, Department of Physics and Joint Center for Quantum Information and Computer Science, University of Maryland, College Park, MD 20742, USA}

\author{Florian Marquardt}
\affiliation{Max Planck Institute for the Science of Light, Staudtstrasse 2, 91058 Erlangen, Germany}
\affiliation{Physics Department, University of Erlangen-Nuremberg, Staudtstrasse 5, 91058 Erlangen, Germany}

\author{Guido Pagano}
\affiliation{Joint Quantum Institute, Department of Physics and Joint Center for Quantum Information and Computer Science, University of Maryland, College Park, MD 20742, USA}
\affiliation{Department of Physics and Astronomy, Rice University, 6100 Main Street, Houston, TX 77005, USA.}

\author{Christopher Monroe}
\affiliation{Joint Quantum Institute, Department of Physics and Joint Center for Quantum Information and Computer Science, University of Maryland, College Park, MD 20742, USA}

\date{\today}

\begin{abstract}
How a closed interacting quantum many-body system relaxes and dephases as a function of time is a fundamental question in thermodynamic and statistical physics.
In this work, we analyse and observe the persistent temporal fluctuations after a quantum quench of a tunable long-range interacting transverse-field Ising Hamiltonian realized with a trapped-ion quantum simulator. We measure the temporal fluctuations in the average magnetization of a finite-size system of spin-$1/2$ particles. We experiment in a regime where the properties of the system are closely related to the integrable Hamiltonian with global spin-spin coupling, which enables analytical predictions even for the long-time non-integrable dynamics. The analytical expression for the temporal fluctuations predicts the exponential suppression of temporal fluctuations with increasing system size. Our measurement data is consistent with our theory predicting the regime of many-body dephasing. 
\end{abstract}

\maketitle

\begin{figure}
  \includegraphics[width=0.99\linewidth]{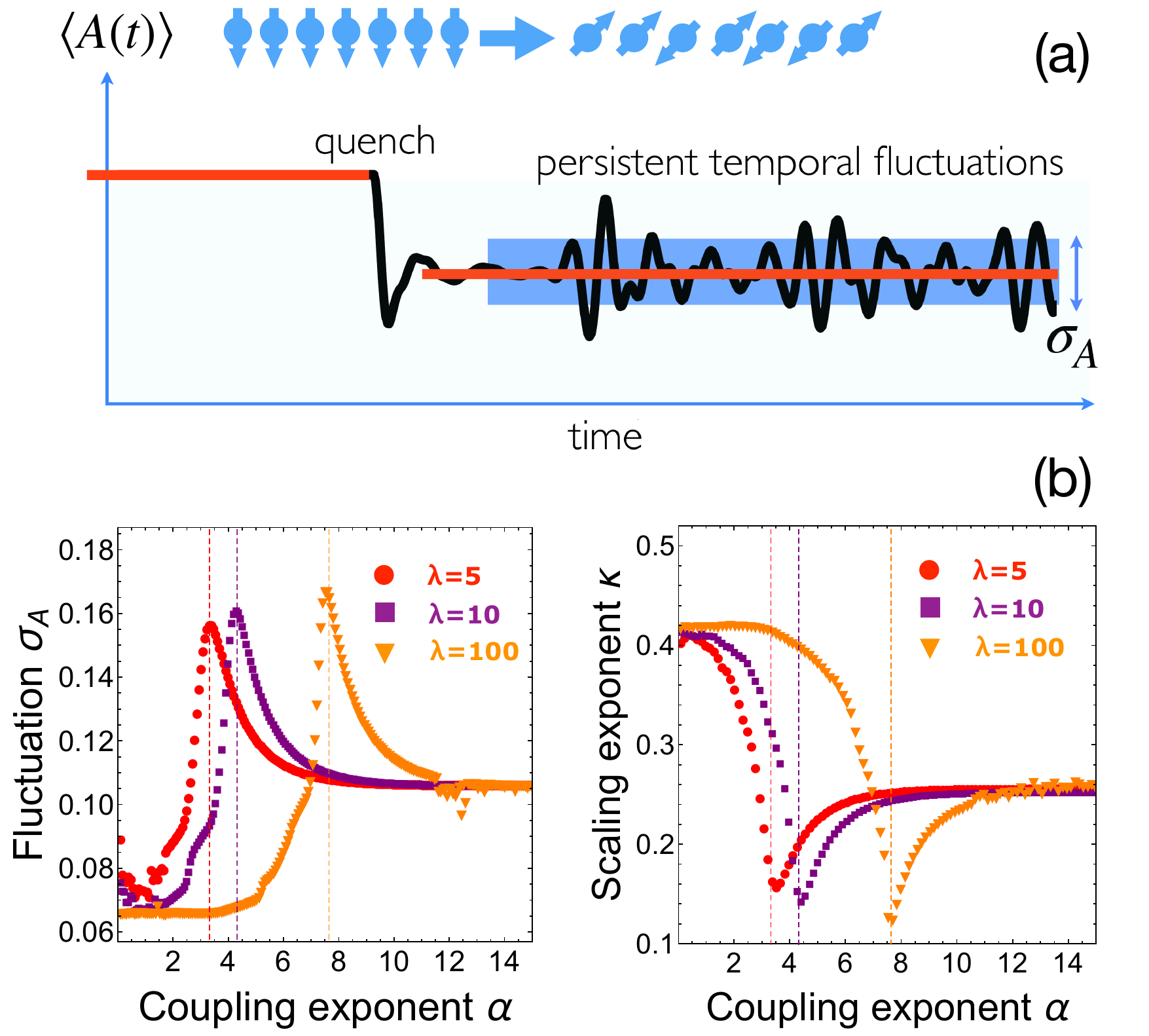}
  \caption{(a) Schematic behaviour of an observable $\langle A(t) \rangle$ after a quench, in a finite-size system. (b) Temporal fluctuation $\sigma_A$ for $N=7$ spins (left) and size scaling exponent $\kappa$ (right) as a function of power-law coupling exponent $\alpha$ for three fixed parameters $\lambda=2J_0/B$. The vertical dashed lines indicate the crossover values of 
  $\alpha^*=\ln(2|\lambda|)/\ln2$ \cite{DynamicalPhaseTransition}. }
  \label{fig_QuantumQuench}
\end{figure}

\textit{Introduction.--} Investigating the relaxation and dephasing dynamics of a closed many-body quantum system is of paramount importance to the study of thermodynamics and statistical physics. Most commonly, this problem is investigated by studying the time evolution of the expectation value of a local observable, e.g., particle density or magnetization, after quenching the system from an initial out-of-equilibrium state \cite{Weiss2006cradle,Polkovnikov2011,Gogolin2016,Bernien2017nature}. For a generic non(near)-integrable system, the expectation value tends to relax to a constant in the thermodynamic limit which can be described by a (pre)thermal state at some temperature depending on the initial state \cite{Rigol2008,Eisert2015,Neuenhahn2012,Deutsch1991,Srednicki1994,Kollath2007,Cramer2008,Trotzky2012,Gring2012,Smith2013,Langen2013,Langen2015,Kaufman2016,Clos2017prl,Neyenhuis2017,Schreiber2015,Smith2016nature,Choi2016}. However, if the system size is finite, there exist persistent temporal fluctuations around the constant average value, as sketched in Fig.~\ref{fig_QuantumQuench}(a). Importantly, these persistent temporal fluctuations in the expectation value after a quench are distinct from the usual fluctuations of observables in equilibrium (where expectation values are constant). Studying these temporal fluctuations represents the next level of the description of quench dynamics going beyond merely looking at long-time observable averages.  

A crucial question for statistical physics is how the temporal fluctuations are  suppressed with increasing system size $N$. In the case of integrable systems mappable to free quasiparticles, it has been found that the variance of temporal fluctuations scales as $1/N$ \cite{Lorenzo2013,Cassidy2011,Gramsch2012}. In the case of generic nonintergrable systems \cite{Reimann2008,Short2012,Lorenzo2014,Herrera2015,Florian2017}, or the integrable systems solvable with the Bethe ansatz (not mappable to noninteracting systems) \cite{Zangara2013}, the temporal fluctuations are exponentially suppressed by the system size due to the highly nondegenerate spectrum. This was first found only numerically. However, in Ref.~\cite{Florian2017}, the authors were able, for the first time, to provide an exact analytical result for the exponential scaling of fluctuations with $N$ spins in a weakly nonintegrable system. In this setting, they identified a general dynamical regime which they termed ``many-body dephasing"\cite{Manybodydephasing}. In the thermalization process, the dephasing mechanism comes from the relaxation of the quasiparticle distribution to thermal equilibrium by quasiparticle scattering described by the Boltzmann equation.
In contrast, many-body dephasing results from lifting of all the exponentially large degeneracies of transition energies in integrable systems while the quasiparticle distribution can remain practically unchanged \cite{Florian2017}. 

Nevertheless, the exponential size scaling due to many-body dephasing in nonintegrable systems has not yet been verified in experiments. Here, we give the first experimental observation of persistent temporal fluctuations after a quantum quench characterized as a function of system size, employing a trapped-ion quantum simulator.
We present a direct measurement of relaxation dynamics in the nonintegrable system by measuring the temporal fluctuations in the average magnetization of a finite-size system of spin-$1/2$ particles. After including the experimental noise in the data analysis, the temporal fluctuations from experimental data are consistent with our numerical simulations and theoretical analysis based on the concept of many-body dephasing.

\textit{Model Hamiltonian.--} The Hamiltonian implemented in this experiment is the long-range transverse-field Ising model, 
\begin{equation}
H=\sum_{i<j} J_{ij}\sigma_i^x\sigma_j^x - \frac{1}{2}B\sum_{i}\sigma_i^z,
\label{Hamiltonian}
\end{equation}
where $J_{ij}\approx J_0/|i-j|^{\alpha}>0$, is a long-range coupling that falls off approximately as a tunable power-law.
The Hamiltonian~(\ref{Hamiltonian}) is implemented using an applied laser field which creates spin-spin interactions through spin-dependent optical dipole forces \cite{SM}.
The spin chain is initialized to the $|\!\!\downarrow\downarrow...\downarrow\rangle_z$ state, then a quench is performed using Hamiltonian~(\ref{Hamiltonian}), and the magnetization along the $z$ axis is measured as a function of time.
The cases of  $\alpha^{-1} = 0$ and $\alpha = 0$ correspond to two integrable limits, i.e., the nearest neighbour coupling and global coupling models respectively.
For a finite $\alpha>0$, Hamiltonian (\ref{Hamiltonian}) is in general nonintegrable.

\textit{Temporal fluctuations.--} In the present experiment, the observable is the magnetization, i.e., $A = N^{-1}\sum_j{\sigma_j^z}$. The temporal average of the variable $\langle  A(t) \rangle$ is calculated as $\overline{\langle  A(t) \rangle} \equiv T^{-1} \int_{t_i}^{t_i+T}\langle A(t)\rangle dt$, where the temporal averaging is restricted within the time window between $t_i$ and $t_i + T$. The variance of temporal fluctuations of $\langle  A(t)\rangle$ is defined via $\sigma_A^2 \equiv \overline{ \big(\langle A(t)\rangle - \overline{\langle A(t)\rangle}\big)^2}$, with $\sigma_A$ the standard deviation.
We use $|\Phi_n\rangle$ ($n=1,2,\cdots,2^N$) to represent the many-body eigenstates of Hamiltonian (\ref{Hamiltonian}) with eigenenergy $E_n$. 
Given the initial state $|\psi(0)\rangle$, the exact time evolution of the observable is
$
\langle A(t)\rangle=\sum_{m,n}\langle\psi(0)|\Phi_m\rangle\langle \Phi_m|A|\Phi_n\rangle\langle\Phi_n|\psi(0)\rangle
e^{i\Delta_{mn}t},
$
where $\Delta_{mn}\equiv E_m-E_n$ is the transition energy between the two energy levels $|\Phi_m\rangle$ and $|\Phi_n\rangle$ ($\hbar=1$). In the long time window limit ($T\rightarrow +\infty$), we have the average 
\bea
\overline{\langle A(t)\rangle}=\sum_{m,n,\Delta_{mn}=0}\langle\psi(0)|\Phi_m\rangle\langle \Phi_m|A|\Phi_n\rangle\langle\Phi_n|\psi(0)\rangle \nonumber 
\eea
and the variance of temporal fluctuation 
\bea\label{VTF}
\sigma^2_A=\sum_{\Delta\neq 0}\Big|\sum_{\Delta_{mn}=\Delta}\langle\psi(0)|\Phi_m\rangle\langle \Phi_m|A|\Phi_n\rangle\langle\Phi_n|\psi(0)\rangle\Big|^2\ \ 
\eea
with  $\Delta$ denoting the set of all the possible values of $\Delta_{mn}$. For the integrable models ($\alpha=0$ or $\alpha^{-1}=0$), there are exponentially many degeneracies with the number of spins for a given transition energy $\Delta_{mn}$, since each many-body eigenstate can be labelled by many independent conserved quantities. However, for a generic nonintegrable model ($\alpha>0$), there are no conserved quantities except the Hamiltonian itself. Thus, it is reasonable to assume that all the degeneracies of transition energies are lifted, making $\Delta_{mn}=0$ only possible for $m=n$ in the nonintegrable model, so Eq.~(\ref{VTF}) simplifies to 
\bea\label{Amn}
\sigma^2_A&=&\sum_{m\neq n}\Big|\langle\psi(0)|\Phi_m\rangle\langle \Phi_m|A|\Phi_n\rangle\langle\Phi_n|\psi(0)\rangle\Big|^2.
\eea
Upon closer analysis, this is the basic reasoning that leads to the exponential suppression of fluctuations with system size \cite{Reimann2008}. However, in general cases, it is impossible to evaluate this expression analytically.

\textit{Theoretical results.--}
We investigate numerically the temporal fluctuation $\sigma_A$ as a function of $\alpha$ for fixed dimensionless parameter $\lambda\equiv 2J_0/B$. We also extract from our numerical simulations the size scaling exponent $\kappa$ from the fit $\sigma_A\propto e^{-\kappa N}$ for $N=3-10$ spins [see Fig.~\ref{fig_QuantumQuench}(b)]. We find two distinct regimes, at small and large $\alpha$, separated by the crossover value of $\alpha^*=\ln(2|\lambda|)/\ln2$ \cite{DynamicalPhaseTransition}. The crossover between those regimes can be understood from the competition between the two terms in Hamiltonian (\ref{Hamiltonian}), i.e., the magnetic field energy $-B\sum_{i}s^z_i$ (where $s^z_i\equiv\frac{1}{2}\sigma_i^z$) and the next-nearest-neighbor (NNN) spin-spin coupling  $2^{-\alpha+2}J_0\sum_{i} s_i^xs_{i+2}^x$, which, for $\alpha>0$, is the leading term responsible for breaking integrability \cite{DynamicalPhaseTransition}. 
In the regime of $\alpha\gg\alpha^*$, by neglecting the NNN (and other long-range) coupling terms, the Hamiltonian is reduced into an integrable model. Adding the NNN coupling terms weakly breaks the integrability and results in many-body dephasing \cite{Florian2017}.
We cannot reach this regime in the experiment since the power-law exponent is $\alpha\approx 0.7$. Therefore this work lies in the opposite regime of $\alpha\ll\alpha^*$, where the long-range coupling terms are dominant over the magnetic field energy, and an analytical prediction can be obtained, as we will show below.

In the global coupling limit ($\alpha=0$), the Hamiltonian 
\bea\label{H0}
H_{\alpha=0}=-B{S}^{z}_{N}+2J_0(S^x_N)^2+NJ_0/2
\eea
is called Lipkin-Meshkov-Glick (LMG) model \cite{Lipkin1965}, which is integrable \cite{Morita2006,Jr2013} since there exist $N$ conserved quantities. For example, $\vec{S}^2_n\equiv S^{x2}_{n}+S^{y2}_{n}+S^{z2}_{n}$ ($n=2,\dots,N$) and the Hamiltonian (\ref{H0}) itself satisfy $[\vec{S}^2_n, H_{\alpha=0}]=0$, where $S^{\beta}_{n}\equiv\sum^{n}_{i=1} \frac{1}{2}\sigma^{\beta}_i$ with $\beta = x,y,z $. In the special case of $\lambda\rightarrow\infty$ ($B\rightarrow0$), we can label each energy level by $|S_1,S_{2},\cdots,S_{N-1},S_N,S^x_N\rangle$ and group all the eigenstates into $N+1$ subspaces according to $S^x_N$. 
In each $S^x_N$-subspace, there are ${N \choose N/2+S^x_N}$ degenerate levels. We define the notation $|\Phi^{\lambda=\infty}_{\frac{N}{2},S^x_N}\rangle$ as the eigenstate with $S_N=N/2$ and spin projection $S^x_N$ at $\lambda=\infty$. 

For finite $\alpha > 0$, since the interaction term in Hamiltonian (\ref{Hamiltonian}) keeps the total spin projection $S^x_N$ unchanged, the eigenstates in different $S^x_N$-subspaces are decoupled. All the degenerate eigenstates in the same $S^x_N$-subspace couple each other resonantly and form new hybridized eigenstates $|\Phi_n\rangle$ appearing in Eq.~(\ref{VTF}).
To estimate $\sigma_A$ in Eq.~(\ref{Amn}), we assume each many-body eigenstate $|\Phi_n\rangle$ to be a superposition of all the ${N \choose N/2+S^x_N}$ levels in the $S^x_N$-subspace with probabilities fluctuating about their uniformly distributed value ${N \choose N/2+S^x_N}^{-1}$.
In the experiment, the pre-quenched spin state is $|\psi(0)\rangle=|\!\downarrow,\downarrow,\cdots,\downarrow\rangle_z$ which only couples the states with total spin $S_N=N/2$. Since $|\Phi^{\lambda=\infty}_{\frac{N}{2},S^x_N}\rangle$ is the only component with total spin $S_N=N/2$ of the many-body $|\Phi_n\rangle$ in the $S^x_N-$subspace, we have
\bea\label{P0}
\big|\langle\psi(0)|\Phi_n\rangle\big|^2\approx{N \choose N/2+S^x_N}^{-1}P^{\lambda=\infty}_{\frac{N}{2},S^x_N}
\eea
with $P^{\lambda=\infty}_{\frac{N}{2},S^x_N}\equiv\big|\langle\psi(0)|\Phi^{\lambda=\infty}_{\frac{N}{2},S^x_N}\rangle\big|^2$. 
Based on this assumption and the eigenstate thermalization hypothesis (ETH) \cite{Deutsch1991,Srednicki1994,Srednicki1999,Beugeling2015,Takashi2018}, we are able to obtain an approximate formula for Eq.~(\ref{Amn}) 
\cite{SM}
\bea\label{sigma2A}
&&\sigma^2_A\approx 2\sum_{S^x_N,S'^x_N}\frac{P^{\lambda=\infty}_{\frac{N}{2},S^x_N}P^{\lambda=\infty}_{\frac{N}{2},S'^x_N}}{{N \choose N/2+S^x_N}+{N \choose N/2+S'^x_N}}\Big|A^{\lambda=\infty}_{S^x_NS'^x_N}\Big|^2
\eea
with the matrix element $A^{\lambda=\infty}_{S^x_NS'^x_N}\equiv\langle \Phi^{\lambda=\infty}_{\frac{N}{2},S^x_N}|A|\Phi^{\lambda=\infty}_{\frac{N}{2},S'^{x}_N}\rangle$.
For large $N$, we have the asymptotic expression that  ${N \choose N/2+S^x_N}\sim 2^N\sqrt{\frac{2}{N\pi}}e^{-2(S^x_N)^2/N}.$
The denominator of Eq.~(\ref{sigma2A}) indicates that $\sigma_A\propto 2^{-N/2}$, predicting the size scaling exponent $\kappa=\ln \sqrt{2}\approx 0.35$. 

Considering \textit{both} $\lambda$ and $\alpha$ finite, the formula (\ref{sigma2A}) holds as long as $\alpha\ll\alpha^*$ but the eigenstate $|\Phi^{\lambda}_{\frac{N}{2},S^x_N}\rangle$ refers to the eigenlevel adiabatically connected to $|\Phi^{\lambda=\infty}_{\frac{N}{2},S^x_N}\rangle$. In general, there is no simple closed form for the eigenstate $|\Phi^{\lambda}_{\frac{N}{2},S^x_N}\rangle$ with a finite $\lambda$. However, Eq.~(\ref{sigma2A}) reduces the calculation of $\sigma_A$ to an $N\times N$ eigenvalue problem which can easily be solved on a computer \cite{Alexander2010}. As we will show further below, the analytical predictions compare well with the experiment (see Fig.~\ref{fig_lnsigmavsN}(a-b) and Ref.~\cite{SM}).

\begin{figure}
  \includegraphics[width=0.99\linewidth]{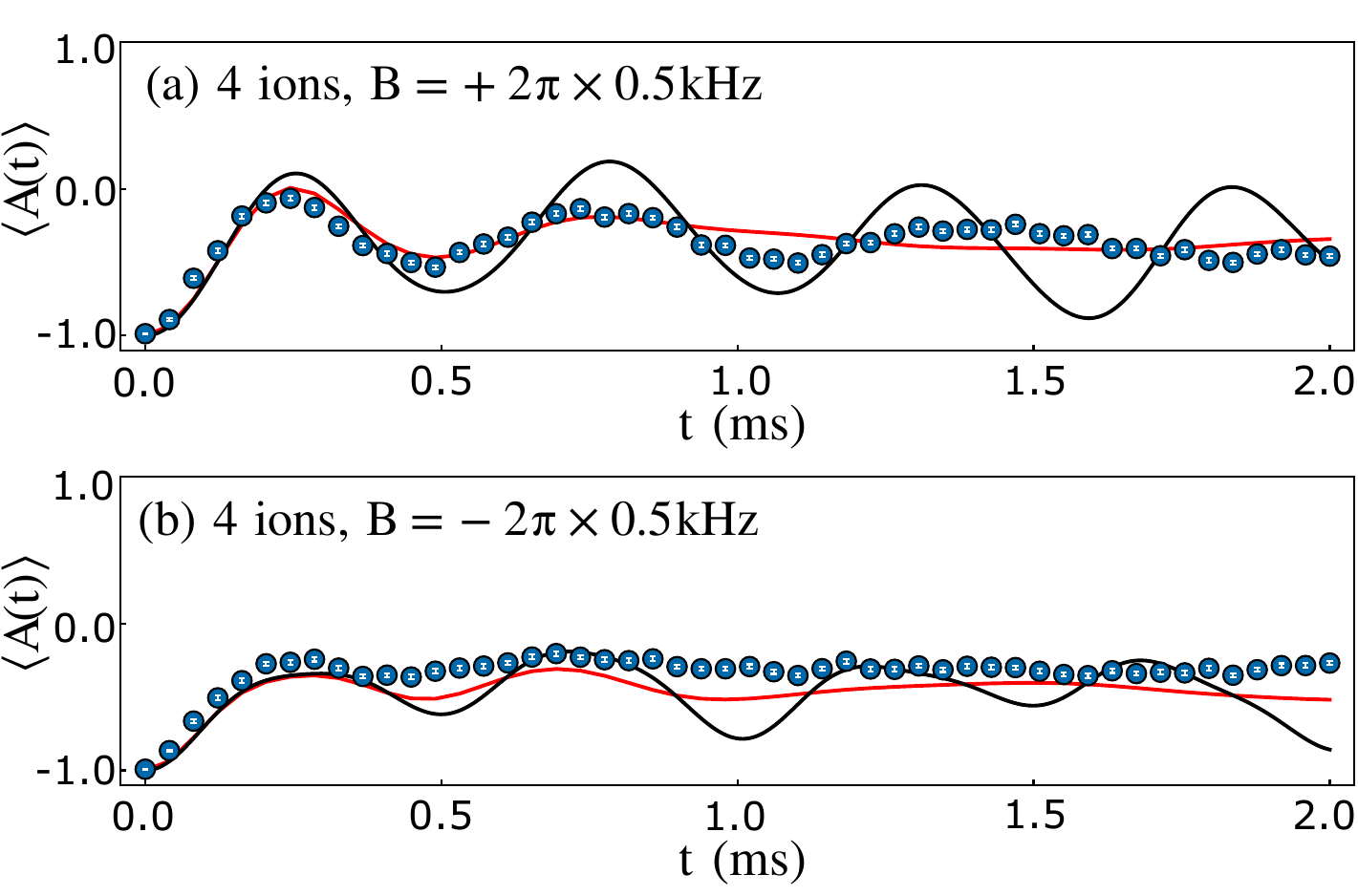}
  \caption{Time evolution of average magnetization, $\langle A\rangle = N^{-1}\sum_j\langle\sigma_j^z\rangle$, over $N=4$ ions out to \SI{2}{\milli\second} for $B=+2\pi\times\SI{0.5}{\kilo\hertz}$ \textbf{(a)} and $B=-2\pi\times\SI{0.5}{\kilo\hertz}$ \textbf{(b)}. Each data point is the average of $4000$ experiments, reported with the respective statistical error (white bars). For both plots: Blue are data points, Black and Red are theoretical results with $(\sigma_{J_0}, \sigma_{B})=0$ and  $(\sigma_{J_0}, \sigma_{B})=2\pi\times(0.1,0.1)\si{\kilo\hertz} $ respectively. Parameters: $J_0=2\pi\times 0.50 \,\si{\kilo\hertz}$, $\alpha=0.71$.}
  \label{magnetization}
\end{figure}

\begin{figure*}
  \includegraphics[width=\textwidth]{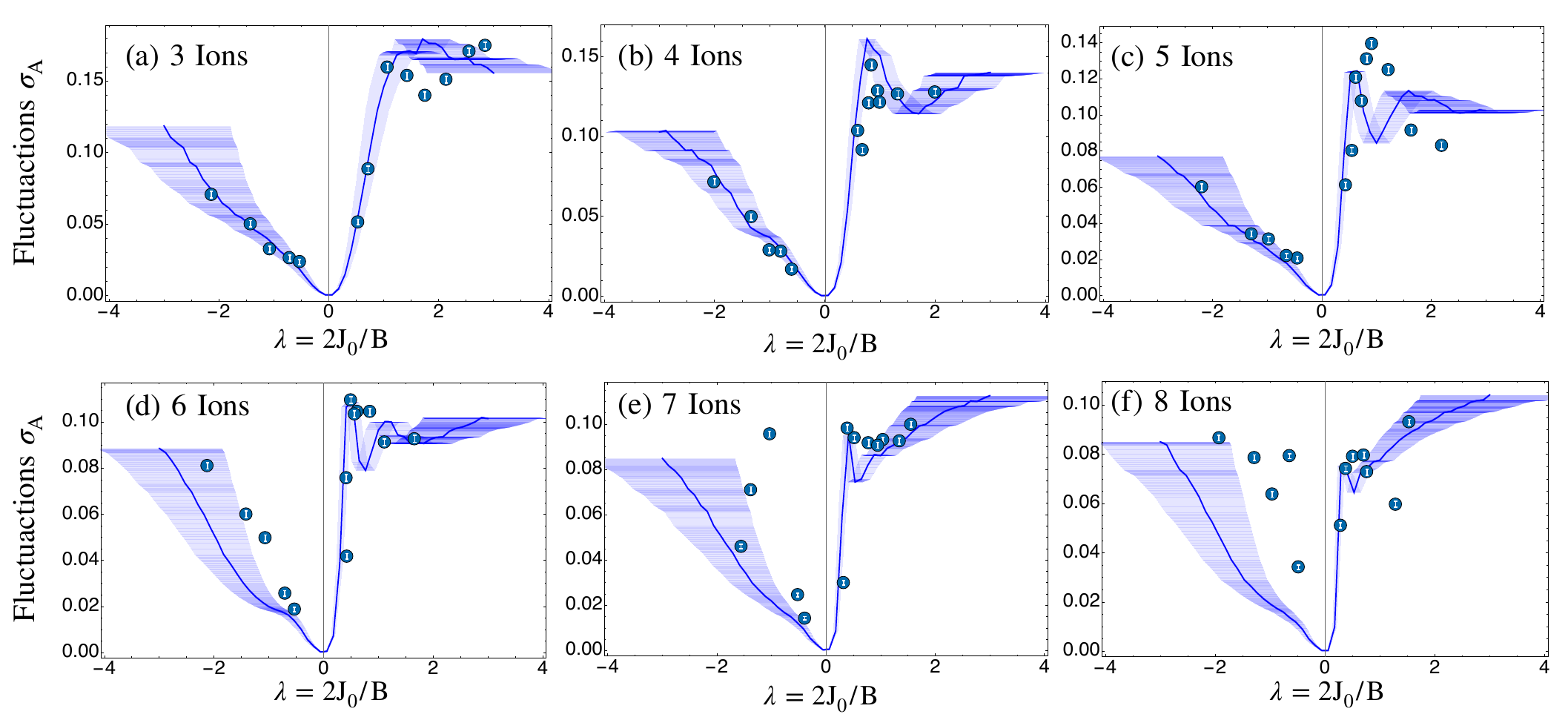}
  \caption{
  Temporal fluctuation $\sigma_A$ as a function of $\lambda = 2J_0/B$ for $N=3-8$ ions from experimental data (blue dots with white error bars) and from numerical simulations (blue curves) with parameters: $(\sigma_{J_0},\sigma_{B}) = 2\pi\times(0.12, 0.12)\,\si{\kilo\hertz}$ for $N=3$, $2\pi\times(0.11, 0.11)\,\si{\kilo\hertz}$ for $N=5$ and $2\pi\times(0.10, 0.10)\,\si{\kilo\hertz}$ for $N=4,6-8$. The blue shade associated with each numerical curve takes account of the experimental uncertainty of $\lambda$.
Experimental parameters: see Table I in Supplementary Material.}
\label{stdev}
\end{figure*}

\textit{Experimental results.--} To perform this experiment, we use a trapped-ion quantum simulator \cite{Pagano2018} where each effective spin $1/2$ particle is encoded in the hyperfine ground state of one $^{171}$Yb$^+$ ion with $|\!\!\uparrow\rangle \equiv ^2$S$_{1/2}|F=1,m_F=0\rangle$ and $|\!\!\downarrow\rangle \equiv ^2$S$_{1/2}|F=0,m_F=0\rangle$.
The Hamiltonian of Eq. (1) is realized by global spin-dependent optical dipole forces from laser beams, which modulate the Coulomb interaction to create an effective Ising coupling between spins \cite{Monroe2019}. The field term is implemented by asymmetrically detuning the two laser beatnotes generating the optical dipole forces \cite{SM}. 

The magnetization fluctuations $\sigma_A$ are characterized by measuring the standard deviation of the average magnetization of the sum of all ions in the chain, i.e., $\langle A\rangle = N^{-1}\sum_j\langle\sigma_j^z\rangle$. This is measured with B-fields ranging from $\pm$ $2\pi\times\SI{0.5}{\kilo\hertz}$ to $2\pi\times\SI{2.0}{\kilo\hertz}$. 
The two plots in Fig.~\ref{magnetization} show the magnetization data measured as a function of time with a $4$-ion chain and $B=\pm 2\pi\times\SI{0.5}{\kilo\hertz}$.
Although the decoherence time in our trapped-ion simulator is long enough to consider $J_0$ and $B$ unchanged within a single time evolution up to $t=2\,\si{\milli\second}$, the values of $J_0$ and $B$ may vary between different time evolutions. We assume the coupling strength and magnetic field in the experiments to be independent and  normally distributed.
Then, the averaged observable $A$ at a fixed time $t$ also needs to be averaged over the experimental values of $J_0$ and $B$, resulting in:
\bea
\langle A(t)\rangle=\langle\langle\psi(t)|A|\psi(t)\rangle\rangle_{J_{0},B}.
\eea
In Fig.~\ref{magnetization}, the red curves are the theory fits by setting $\sigma_{J_0}$ and $\sigma_{B}$ both to approximately $2\pi\times\SI{0.1}{\kilo\hertz}$. To fit the experimental data, we use the gradient descent method to search for the optimal values of $\sigma_{J_0}$ and $\sigma_{B}$, which happen to be roughly equal. Therefore, we set $\sigma_{J_0}$ and $\sigma_{B}$ to be the same values for simplicity.

In general, with a positive B-field, we observe more significant oscillations than when using a negative B-field. This can be understood by analyzing the overlap between the pre-quench state and the post-quench energy eigenstates (obtained for the post-quench $J_0$ and $B$ values). For the system parameters given in Fig.~\ref{magnetization}, the structure of the post-quench spectrum is such that at high energies there is a non-vanishing energy gap in the thermodynamic limit. Conversely, in the low energy sector of the spectrum the level spacing decreases with system size and the gap vanishes in the thermodynamic limit. For the positive B-field, the pre-quench state is the superposition of several of the highest excited states of the spectrum and the energy gap leads to more persistent oscillations. For the negative B-field, the pre-quench state is very close to the ground state of the spectrum \cite{Jurcevic2017}, suppressing the oscillations \cite{SM}.

We plot the standard deviation of the average magnetization $\sigma_A$ as a function of $\lambda = 2J_0/B$ for fixed $N$ in Fig.~\ref{stdev}.
The data for $N=3$ to $N=6$ agree with the theoretical prediction. The $N=7$ data largely agrees with theory excluding the two outlying points at negative $\lambda$ values. For $N=8$, the data points tend to gather around the 0.07 level indicating that the measurement noise in this case obscures the measured fluctuations.
In these plots, the values near $\lambda=0$ were not taken because when $B\gg J_0$ the ions are predominantly acting paramagnetically.
In this regime, fluctuations are expected to be very small and well below the noise floor of this experiment.
The shape of the data is asymmetric with a pronounced slope at $2J_0/B = 1/2$. This point marks the ferromagnetic (FM) to paramagnetic (PM) phase transition of the ion chain. The fluctuations are enhanced here as this is an unstable point for the system.
In contrast, the antiferromagnetic (AFM) to PM transition \cite{PhaseTransition} for $\lambda<0$ is not as pronounced.

\begin{figure}
\includegraphics[width=0.99\linewidth]{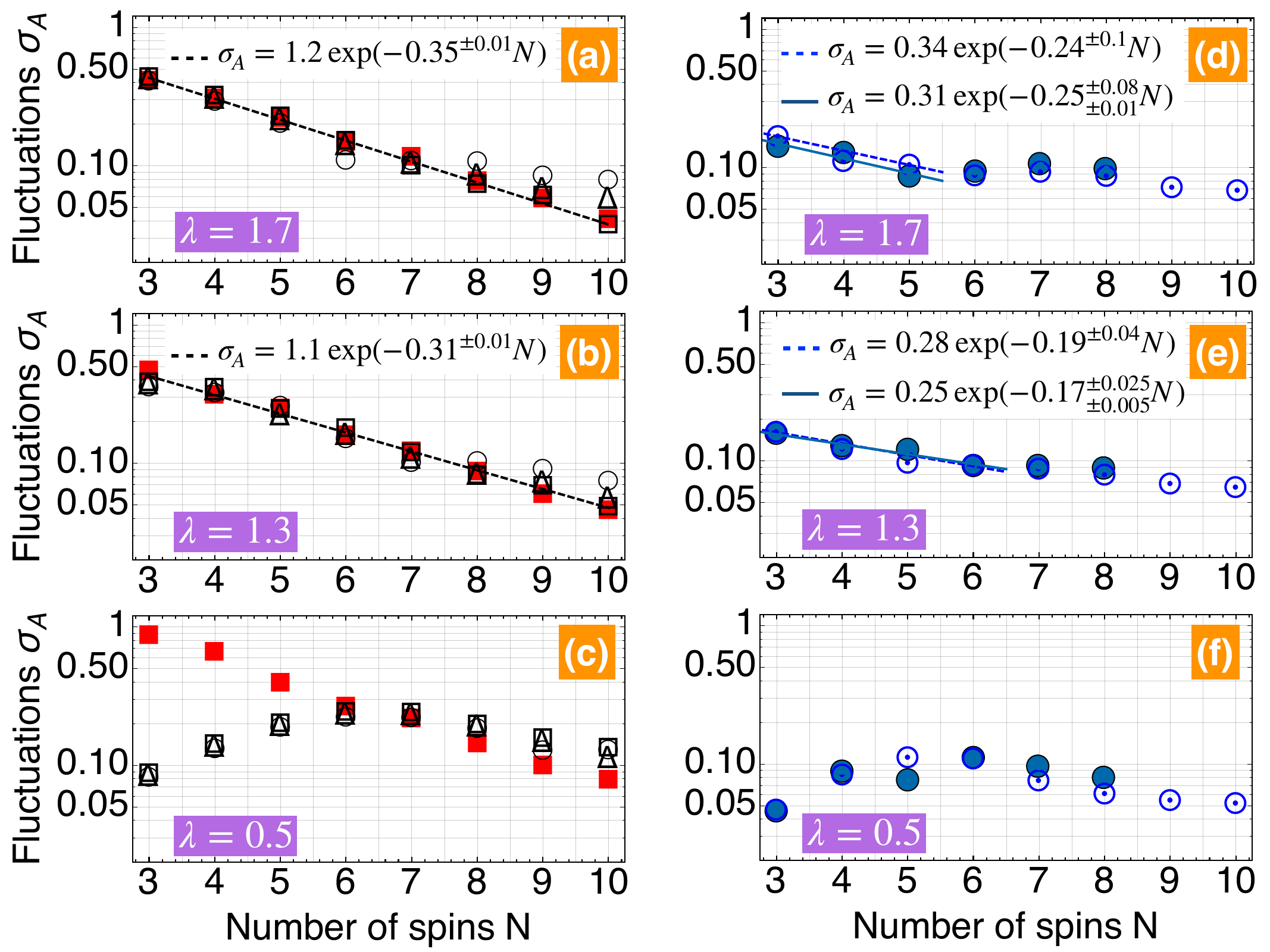}
\caption{\label{fig_lnsigmavsN}
{Logplot of temporal fluctuation $\sigma_A$ versus number of spins $N$ for different $\lambda$ values:} (a)-(c) Analytical results (red squares) vs numerical results with different time windows $Jt\in [0,2\pi]$ (circles), $Jt\in [0,5\pi]$ (triangles) and $Jt\in [0,\infty]$ (empty squares) for $\alpha=0.7$.
(d-f) Experimental results (blue dots) vs
numerical results (circled dots) taken from Fig.~\ref{stdev}. The dashed and solid lines are the fits.}
\end{figure}

\textit{System size scaling.--} The temporal fluctuation variance $\sigma^2_A$ given by Eq.~(\ref{Amn}) is obtained by averaging over an \textit{infinite} time window $J_0t\in [0,+\infty]$. However, in the experiment, we can only average over a \textit{finite} time window up to $t\sim 2.0\, \si{\milli\second}$ ( i.e., 3 or 4 oscillations depending on the value of $\lambda$), as the long-time fluctuations are suppressed by the noise in the parameters $J_0$ and $B$. 
 In Figs.~\ref{fig_lnsigmavsN}(a)-(c), we compare the analytical results given by Eq.~(\ref{sigma2A}) to the numerical results with different averaging time windows. The short-time-window averaging only makes sense for small system size as
larger system sizes result in smaller level splittings and makes the period of temporal fluctuations longer. The fits to the infinite-time-window averaging show that the system size scaling exponent is $\kappa\approx 0.31$ $(\kappa\approx 0.35)$ for $\lambda=1.3$ ($\lambda=1.7$), which is close to the theoretical prediction $\kappa=\ln\sqrt{2}$ for $N\gg 1$ \cite{SM}.

In Figs.~\ref{fig_lnsigmavsN}(d)-(f), we compare the experimental data with the numerical results taken from Fig.~\ref{stdev} for different $\lambda$ values.
The fits to the experimental data and numerical results for $N=3-5$ ($\lambda=1.7$) or $N=3-6$ ($\lambda=1.3$) show good agreement. 
For $\lambda=0.5$, our analytical expression breaks down as the system is in the crossover regime, but the experimental data still confirm the the numerical simulations as shown in Fig.~\ref{fig_lnsigmavsN}(f). 
For $\lambda=1.7$, the fit to the experimental data gives the system size scaling exponent $\kappa=0.25^{\pm 0.08}_{\pm 0.01}$, where the superscript is the uncertainty from the least square fitting and the subscript is the uncertainty from the statistical errors in the experiments \cite{SM}.
We finally note that an exponential fit of data generated from single-particle dephasing ($\sigma_A\propto 1/\sqrt{N}$) \cite{Florian2017} for $N = 3-5$
would lead to an exponent $\kappa\sim 0.13$, which is expected to be even  further suppressed by the noise in $J_0$ an $B$. A detailed statistical analysis is presented in the supplementary information. \cite{SM}.

\textit{Summary.--} Using a trapped-ion quantum simulator, we have presented the first experimental observation of persistent temporal fluctuations after a quantum quench with a long-range interacting transverse-field Ising model. We characterized how the fluctuations in the average magnetization of the spin chain depend on the transverse field and the spin-spin interactions.
Numerical simulations
compared with experiment show that, as a function of
system size $N$, the exponential suppression of temporal fluctuations matches well with the theoretical prediction.

\begin{acknowledgments}
{\it Acknowledgements}.--- 
This work is supported by the NSF PFCQC STAQ program, the AFOSR MURIs on Quantum Measurement/Verification, the ARO MURI on Modular Quantum Systems, the DARPA DRINQS program, the DOE BES award de-sc0019449, the DOE HEP award de-sc0019380, and the NSF Physics Frontier Center at JQI.
\end{acknowledgments}


\bibliographystyle{plain}
\bibliography{references}

\begin{thebibliography}{9}

\bibitem{Weiss2006cradle}
T. Kinoshita, T. Wenger, and D. Weiss, Nature {\bf 440}, 900 (2006).
\bibitem{Polkovnikov2011}
A. Polkovnikov, K. Sengupta, A. Silva, and M.
Vengalattore, Rev. Mod. Phys. {\bf 83}, 863 (2011).
\bibitem{Gogolin2016}
Christian Gogolin and Jens Eisert, Rep. Prog. Phys. {\bf 79} 056001 (2016).
\bibitem{Bernien2017nature}
H. Bernien, S. Schwartz, A. Keesling, \textit{et al.}, Nature {\bf 551}, 579 (2017). 


\bibitem{Rigol2008}
M. Rigol, V. Dunjko, and M. Olshanii, Nature (London)
{\bf 452}, 854 (2008).
\bibitem{Eisert2015}
J. Eisert, M. Friesdorf, and C. Gogolin, Nat. Phys. {\bf 11}, 124
(2015).
\bibitem{Neuenhahn2012}
C. Neuenhahn and F. Marquardt, Phys. Rev. E {\bf 85}, 060101(R)
(2012).
\bibitem{Deutsch1991}
J. M. Deutsch, Phys. Rev. A {\bf 43}, 2046 (1991).
\bibitem{Srednicki1994}
M. Srednicki, Phys. Rev. E {\bf 50}, 888 (1994).
\bibitem{Kollath2007}
C. Kollath, A. M. L\"auchli, and E. Altman, Phys. Rev. Lett.
{\bf 98}, 180601 (2007).
\bibitem{Cramer2008}
M. Cramer, C. M. Dawson, J. Eisert, and T. J. Osborne,
Phys. Rev. Lett. {\bf 100}, 030602 (2008).

\bibitem{Trotzky2012}
S. Trotzky, Y. Chen, A. Flesch, \textit{et al.}, Nat. Phys. {\bf 8}, 325(2012).
\bibitem{Gring2012}
M. Gring, \textit{et al.}, Science {\bf 337}, 1318(2012).
\bibitem{Smith2013}
D. A. Smith \textit{et al.}, New J. Phys. {\bf 15}, 075011(2013)
\bibitem{Langen2013}
T. Langen, R. Geiger, M. Kuhnert \textit{et al.}, Nat. Phys. {\bf 9}, 640(2013). 
\bibitem{Langen2015}
T. Langen \textit{et al.}, Science {\bf 348}, 207(2015).
\bibitem{Kaufman2016}
A. M. Kaufman \text{et al.}, Science {\bf 353}, 794(2016).
\bibitem{Clos2017prl}
G. Clos, D. Porras, U. Warring, and T. Schaetz, Phys. Rev. Lett. {\bf 117}, 170401 (2016).
\bibitem{Neyenhuis2017}
B. Neyenhuis \textit{et al.}, Science Advances {\bf 3}, e1700672(2017).
\bibitem{Schreiber2015}
M. Schreiber \textit{et al.}, Science {\bf 349}, 842(2015)
\bibitem{Smith2016nature}
J. Smith, A. Lee, P. Richerme \textit{et al.}, Nat. Phys. {\bf 12}, 907(2016).
\bibitem{Choi2016}
Jae-yoon Choi \textit{et al.}, Science {\bf 352}, 1547(2016)


\bibitem{Lorenzo2013}
Lorenzo Campos Venuti and Paolo Zanardi,
Phys. Rev. E {\bf 87}, 012106 (2013)
\bibitem{Cassidy2011}
Amy C. Cassidy, Charles W. Clark, and Marcos Rigol,
Phys. Rev. Lett. {\bf 106}, 140405 (2011)
\bibitem{Gramsch2012}
Christian Gramsch and Marcos Rigol,
Phys. Rev. A {\bf 86}, 053615 (2012)

\bibitem{Reimann2008}
Peter Reimann, Phys. Rev. Lett. {\bf 101}, 190403 (2008).
\bibitem{Short2012}
A. J. Short and T. C. Farrelly, New J. Phys. {\bf 14}, 013063
(2012).
\bibitem{Lorenzo2014}
Lorenzo Campos Venuti and Paolo Zanardi,
Phys. Rev. E {\bf 89}, 022101 (2014).
\bibitem{Herrera2015}
E. J. Torres-Herrera, D. Kollmar, and L. F. Santos, Phys. Scr.
2015, 014018 (2015).
\bibitem{Florian2017}
  Kiendl, T., Marquardt, F., Phys. Rev. Lett. {\bf 118}, 130601 (2017).
  
\bibitem{Zangara2013} 
P. R. Zangara, A. D. Dente, E. J. Torres-Herrera, H. M.
Pastawski, A. Iucci, and L. F. Santos, Phys. Rev. E {\bf 88}, 032913 (2013). 

\bibitem{Pagano2018} G. Pagano, \emph{et al.}, Quantum Science and Technology {\bf 4}, 014004 (2018)

\bibitem{Manybodydephasing} 
In Ref.~\cite{Florian2017}, the authors used the name ``many-particle dephasing" since their model can be reduced into weakly-interacting quasi-particles. But for the model considered in our work, there is no clear quasi-particle picture. Thus, we rephrase the name as ``many-body dephasing" here.

\bibitem{SM} 
See the Supplemental Material for details, which includes Refs.~\cite{smWigner,smDyson,smRMT,smRigol,smKota,smBerry,Srednicki1994,Deutsch1991,Srednicki1999,Beugeling2015,Takashi2018,Florian2017,Pagano2018,smMolmer,smKim2009,smAaronThesis,smOlmschenk2007,smWang2012,smYukaiThesis,smSTDEVCI}.

\bibitem{DynamicalPhaseTransition}
 This crossover may be related to the phenomenon of dynamical phase transition  \cite{Heyl2013,Heyl2014,Monroe2017,Halimeh2017prb1,Halimeh2017prb2,Halimeh2017pre,Heyl2018}. But it is not the focus of the present work, so we leave its study to future investigations. The crossover value of $\alpha^*=\ln(2|\lambda|)/\ln2$ can be obtained by comparing  the magnetic field energy and the next-nearest-neighbor (NNN) coupling term, i.e., $|2^{-\alpha^*+2}J_0|=|-B|$.
%
Note that this crossover is different from the well-known paramagnetic to antiferromagnetic phase transition discussed in Ref.~\cite{PhaseTransition}.

\bibitem{Srednicki1999}
Mark Srednicki, J. Phys. A: Math. Gen. {\bf 32}, 1163 (1999)
\bibitem{Beugeling2015}
Wouter Beugeling, Roderich Moessner, and Masudul Haque
Phys. Rev. E {\bf 91}, 012144 (2015).
\bibitem{Takashi2018}
Takashi Mori et al, J. Phys. B: At. Mol. Opt. Phys. {\bf 51}, 112001 (2018).


\bibitem{Heyl2013}
M. Heyl, A. Polkovnikov, and S. Kehrein,
Phys. Rev. Lett. {\bf 110}, 135704 (2013).
\bibitem{Heyl2014}
M. Heyl, Phys. Rev. Lett. {\bf 113}, 205701 (2014).
\bibitem{Monroe2017}
J. Zhang \textit{et al}, Nature {\bf 551}, 601 (2017).
\bibitem{Halimeh2017prb1}
J. C. Halimeh, V. Zauner-Stauber, I. P. McCulloch, I. de Vega, U. Schollw\"ock, and M. Kastner, Phys. Rev. B {\bf 95}, 024302(2017).
\bibitem{Halimeh2017prb2}
J. C. Halimeh and V. Zauner-Stauber, Phys. Rev. B {\bf 96}, 134427(2017).
\bibitem{Halimeh2017pre}
V. Zauner-Stauber and J. C. Halimeh, Phys. Rev. E {\bf 96}, 062118 (2017).
\bibitem{Heyl2018}
B. \^Zunkovi\^c, M. Heyl, M. Knap, and A. Silva,
Phys. Rev. Lett. {\bf 120}, 130601 (2018).


\bibitem{Lipkin1965}
H. Lipkin, N. Meshkov, and A. Glick, Nucl. Phys. {\bf 62}, 188 (1965).
\bibitem{Morita2006}
H. Morita, H. Ohnishi, J. da Provid\^encia and S. Nishiyama, Nucl. Phys. B {\bf 737}, 337 (2006).
\bibitem{Jr2013}
W. M. Jr, S. Post and P. Winternitz, J. Phys. A: Math. Theor. {\bf 46}, 423001 (2013).

\bibitem{Alexander2010}
J. A. Alexander, P. Reinhard and E. Suraud, \textit{Simple Models of Many-Fermion Systems} (Spring-Verlag Berlin Heidelberg 2010, Page 171).

 \bibitem{Monroe2019}
C. Monroe
 \textit{et al}, arXiv:1912.07845 (2019).

 \bibitem{Jurcevic2017}
P. Jurcevic
 \textit{et al}, Phys. Rev. Lett. {\bf 119}, 080501 (2017).

 \bibitem{PhaseTransition}
T. Koffel, M. Lewenstein, and L. Tagliacozzo, Phys. Rev. Lett. {\bf 109}, 267203 (2012).


\bibitem{smWigner}
E. P. Wigner, Ann. Math. {\bf 62}, 548 (1955); E. P. Wigner, Ann. Math. {\bf 65}, 203 (1957); E. P. Wigner, Ann. Math. {\bf 67}, 325 (1958).
\bibitem{smDyson}
F. J. Dyson, J. Math. Phys. {\bf 3}, 140 (1962).
\bibitem{smRMT}
M. L. Mehta, {\it Random Matrices}, Elsevier/Academic Press, Amsterdam, 2004.
\bibitem{smRigol}
L. F. Santos and M. Rigol, Phys. Rev. E {\bf 81}, 036206 (2010).
\bibitem{smKota}
V. K. B. Kota, Lecture Notes Phys. 884 (2014), page 31.
\bibitem{smBerry}
M. V. Berry, J. Phys. A {\bf 10}, 2083 (1977).

\bibitem{smMolmer}
K. Molmer and K. Sorensen
\textit{et al.}, Phys. Rev. Lett. {\bf 82}, 1835 (1999)

\bibitem{smKim2009}
K. Kim
\textit{et al.}, Phys. Rev. Lett. {\bf 103}, 120502 (2009).

\bibitem{smAaronThesis}
A. C. Lee,
\textit{Engineering a Quantum Many-Body Hamiltonian with Trapped Ions}, Aaron Lee Thesis (2016)
 
\bibitem{smOlmschenk2007}
Olmschenk, S., \textit{et al.}, Phys. Rev. A, {\bf 76}, 052314 (2007).

\bibitem{smWang2012} 
C. -C. Joseph Wang, J. K. Freericks,
\emph{Phys. Rev. A} \textbf{86}, 032329 (2012).

\bibitem{smYukaiThesis} 
Y. Wu, thesis, University of Michigan (2019).

\bibitem{smSTDEVCI}
D. J. Sheskin,
\textit{Handbook of Parametric and Nonparametric Statistical Procedures}, 4th Edition, IBSN:1584888148.


\end{thebibliography}

\begin{thebibliography}{9}

\bibitem{Wigner}
E.P. Wigner, Ann. Math. {\bf 62}, 548 (1955); E.P. Wigner, Ann. Math. {\bf 65}, 203 (1957); E.P. Wigner, Ann. Math. {\bf 67}, 325 (1958).
\bibitem{Dyson}
F.J. Dyson, J. Math. Phys. {\bf 3}, 140 (1962).
\bibitem{RMT}
M.L. Mehta, {\it Random Matrices}, Elsevier/Academic Press, Amsterdam, 2004.
\bibitem{Rigol}
L.F. Santos and M. Rigol, Phys. Rev. E {\bf 81}, 036206 (2010).
\bibitem{Kota}
V.K.B. Kota, Lecture Notes Phys. 884 (2014), page 31.
\bibitem{Berry}
M.V. Berry, J. Phys. A {\bf 10}, 2083 (1977).
\bibitem{Srednicki1994}
M. Srednicki, Phys. Rev. E {\bf 50}, 888 (1994).

\bibitem{Deutsch1991}
J. M. Deutsch, Phys. Rev. A {\bf 43}, 2046 (1991).
\bibitem{Srednicki1999}
Mark Srednicki, J. Phys. A: Math. Gen. {\bf 32}, 1163 (1999)
\bibitem{Beugeling2015}
Wouter Beugeling, Roderich Moessner, and Masudul Haque
Phys. Rev. E {\bf 91}, 012144 (2015).
\bibitem{Takashi2018}
Takashi Mori \textit{et al.}, J. Phys. B: At. Mol. Opt. Phys. {\bf 51}, 112001 (2018).
\bibitem{Florian2017}
  Kiendl, T., Marquardt, F., Phys. Rev. Lett. {\bf 118}, 130601 (2017).

\bibitem{CryoTechnical}
G. Pagano
\textit{et al.}, Quantum Science and Technology {\bf 4}, 014004 (2019).

\bibitem{Molmer}
K. Molmer and K. Sorensen
\textit{et al.}, Phys. Rev. Lett. {\bf 82}, 1835 (1999)

\bibitem{Kim2009}
K. Kim
\textit{et al.}, Phys. Rev. Lett. {\bf 103}, 120502 (2009).

\bibitem{AaronThesis}
A. C. Lee,
\textit{Engineering a Quantum Many-Body Hamiltonian with Trapped Ions}, Aaron Lee Thesis (2016)
 
\bibitem{Olmschenk2007}
Olmschenk, S., \textit{et al.}, Phys. Rev. A, {\bf 76}, 052314 (2007).

\bibitem{Wang2012} C. -C. Joseph Wang, J. K. Freericks,
\emph{Phys. Rev. A} \textbf{86}, 032329 (2012).

\bibitem{YukaiThesis} Y. Wu, thesis, University of Michigan (2019).

\bibitem{STDEVCI}
D. J. Sheskin,
\textit{Handbook of Parametric and Nonparametric Statistical Procedures}, 4th Edition, IBSN:1584888148.

\bibitem{LSF}
https://mathworld.wolfram.com/LeastSquaresFitting.html

\end{thebibliography}

\newpage\ 

\newpage

\onecolumngrid
\section{Supplemental Material for \\ \textit{Many-Body Dephasing in a Trapped-Ion Quantum Simulator} }
\twocolumngrid

\section{{I. Temporal fluctuations}}\label{TF}
\begin{figure*}
  \includegraphics[width=1\linewidth]{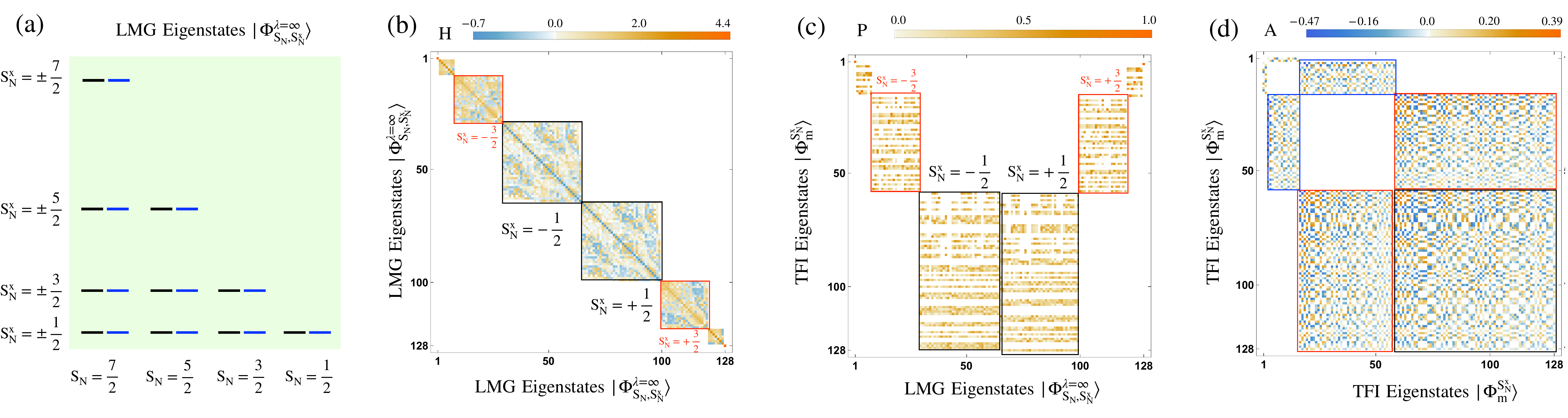}
  \caption{ Transverse-field Ising (TFI) model in the parameter limit $\lambda\rightarrow\infty$ (or $B= 0$).   (a) Energy level structure of $N=7$ ions for the power-law coupling exponent $\alpha=0$, i.e., LMG model. $S_N$ is the total spin quantum number of $N$ ions and $S^x_N$ is the total spin component along the $x$-axis. Each row contains two degenerate subspaces with opposite $S^x_N$ (black and blue lines) due to $E\propto (S^x_N)^2$ for the LMG model with $\lambda=\infty$. (b) Matrix elements of TFI Hamiltonian ($\alpha=0.2$, $N=7$ ions) in the eigenstate basis of LMG model ($\alpha=0$, $N=7$ ions). The Hamiltonian shows a diagonal block-matrix structure since the finite-$\alpha$ coupling term only hybridizes the levels inside each $S^{x}_N$ subspace of LMG model. While the diagonal matrix elements in each block (symmetry sector) are nearly a constant, the off-diagonal elements in each symmetry sector behave like random numbers.  (c) The probability of TFI Hamiltonian ($\alpha=0.2$, $N=7$) eigenstates (vertical axis) projected on the LMG model eigenstates (horizontal axis). Every TFI eigenstate is basically smeared over the degenerate energy levels in each $S^x_N$ subspace of LMG model. (d) Matrix elements of observable $A=N^{-1}\sum_j\sigma^z_j$ in the eigenstate basis of TFI Hamiltonian ($\alpha=0.2$, $N=7$). The block-matrix structure comes from the fact that  operator $A=N^{-1}\sum_j\sigma^z_j$ only couples nearest-neighbouring $S^x_N$ subspaces of TFI Hamiltonian, e.g., the two blocks with black (red, blue) borders are from the coupling between $S^x_N=-\frac 1 2$ and $S^x_N=+\frac 1 2$ subspaces ($S^x_N=\pm \frac 1 2$ and $S^x_N=\pm \frac 3 2$ subspaces, $S^x_N=\pm \frac 3 2$ and $S^x_N=\pm \frac 5 2$ subspaces). The random-like matrix elements of these blocks can be described by ETH (\ref{ETH}), but the variances $\overline{A^2}$ in different blocks are not the same, which can be distinguished by the colour intensity in the plots. } 
  \label{fig_LevelStructure}
\end{figure*}

Here, our aim is to calculate the variance of temporal fluctuation given by
\bea\label{sigma2Asm}
\sigma^2_A&=&\sum_{m\neq n}\Big|\langle\psi(0)|\Phi_m\rangle\langle \Phi_m|A|\Phi_n\rangle\langle\Phi_n|\psi(0)\rangle\Big|^2.
\eea
In Fig.~\ref{fig_LevelStructure}(a), we show the energy level structure for the LMG model for $N=7$ ions in the limit $\lambda\rightarrow \infty$, where the eigenstates are labeled by $|\Phi^{\lambda=\infty}_{S_N,S^x_N}\rangle$ with $S_N$ the total quantum spin number and $S^x_N$ the total spin component along $x-$direction. A finite-$\alpha$ coupling term in the transverse-field Ising (TFI) model only hybridises the energy levels inside each $S^x_N$ subspace. Therefore, we classify all the eigenstates of TFI model into different $S^x_N$ subspaces and use $|\Phi^{S^x_N}_m\rangle$ to represent the TFI eigenstate in each $S^x_N$ subspace. In this way, the above expression (\ref{sigma2Asm}) can be written alternatively
\bea\label{sigma2A2}
\sigma^2_A&=&\sum_{S^x_N,S'^x_N}\sum_{m\neq m'}\Big|\langle\psi(0)|\Phi^{S^x_N}_m\rangle\langle \Phi^{S^x_N}_m|A|\Phi^{S'^x_N}_{m'}\rangle\langle\Phi^{S'^x_N}_{m'}|\psi(0)\rangle\Big|^2.\nonumber\\
\eea
Here, $m(m')$ are the indices of the eigenstates in the $S^x_N(S'^x_N)$ subspace by sorting their energy levels.

In the experiment, the pre-quenched spin state $|\psi(0)\rangle=|\!\downarrow,\downarrow,\cdots,\downarrow\rangle_z$ only couples the states with total spin $S_N=N/2$. Thus, $|\Phi^{\lambda=\infty}_{\frac{N}{2},S^x_N}\rangle$ is the only component coupled to $|\psi(0)\rangle$ in the $S^x_N$ subspace. Therefore, we have
\bea\label{P0}
\big|\langle\psi(0)|\Phi^{S^x_N}_m\rangle\big|^2=\big|\langle\psi(0)|\Phi^{\lambda=\infty}_{\frac{N}{2},S^x_N}\rangle\big|^2\big|\langle\Phi^{\lambda=\infty}_{\frac{N}{2},S^x_N}|\Phi^{S^x_N}_m\rangle\big|^2.
\eea
Clearly, the quantity $\big|\langle\Phi^{\lambda=\infty}_{\frac{N}{2},S^x_N}|\Phi^{S^x_N}_m\rangle\big|^2$ is the probability of TFI Hamiltonian eigenstate $|\Phi^{S^x_N}_m\rangle$ projected on the LMG basis $|\Phi^{\lambda=\infty}_{\frac{N}{2},S^x_N}\rangle$. The TFI Hamiltonian is non-intergrable for $\alpha\neq 0$. In Fig.~\ref{fig_LevelStructure}(b), we plot the matrix elements of TFI Hamiltonian ($\alpha=0.2$, $N=7$ ions) in the LMG basis $|\Phi^{\lambda=\infty}_{S_N,S^x_N}\rangle$. In general, as pointed out by Wigner \cite{Wigner}, it is hopeless to try to predict the exact eigenlevels and eigenstates of a complex non-intergrable quantum Hamiltonian. Instead, one can use a {\it random matrix} to model the non-intergrable Hamiltonian in a non fine-tuned basis (e.g., it is a fine-tuning of the basis to write the Hamiltonian in its own basis). Wigner's seminal works, followed by Dyson \cite{Dyson}, are now known as random matrix theory (RMT) \cite{RMT}. In fact, in real systems, the Hamiltonian matrix has more structure than a fully Gaussian random matrix. For instance, one should account for additional symmetries of Hamiltonian and look at symmetry sectors as they are decoupled \cite{Rigol}. As shown by Fig.~\ref{fig_LevelStructure}(b), our TFI Hamiltonian ($\alpha=0.2$, $N=7$) shows a diagonal block-matrix structure since the finite-$\alpha$ coupling term hybridizes the levels inside each $S^{x}_N$ subspace of LMG model. While the diagonal matrix elements in each block are nearly a constant, the off-diagonal elements behave like random numbers. In Fig.~\ref{fig_LevelStructure}(c), we plot the probability projection of TFI Hamiltonian ($\alpha=0.2$, $N=7$) eigenstates $|\Phi^{S^x_N}_m\rangle$ (vertical axis) on the LMG model eigenstates $|\Phi^{\lambda=\infty}_{S_N,S^x_N}\rangle$ (horizontal axis). Every TFI eigenstate $|\Phi^{S^x_N}_m\rangle$ is basically smeared over the degenerate energy levels in the corresponding $S^x_N$ subspace of LMG model. As predicted by RMT \cite{Kota} and also conjectured by Berry \cite{Berry,Srednicki1994}, the superposition coefficient of a non-integrable Hamiltonian's eigenstate on a non-fine tuned basis is a Gaussian random variable with averaged probability of $1/D$, where $D$ is the dimension of the relevant Hilbert space. In each $S^x_N$ symmetry sector of TFI Hamiltonian, we make the assumption that the superposition coefficients $\langle\Phi^{\lambda=\infty}_{\frac{N}{2},S^x_N}|\Phi^{S^x_N}_m\rangle$ are random numbers with the averaged $\overline{\big|\langle\Phi^{\lambda=\infty}_{\frac{N}{2},S^x_N}|\Phi^{S^x_N}_m\rangle\big|^2}=1/D$, where $D={N \choose N/2+S^x_N}$ is the dimension of $S^x_N$ subspace. Therefore, we take $\big|\langle\Phi^{\lambda=\infty}_{\frac{N}{2},S^x_N}|\Phi^{S^x_N}_m\rangle\big|^2\approx {N \choose N/2+S^x_N}^{-1}$ as an approximation and simplify Eq.~(\ref{P0}) into 
\bea\label{sigma2App}
\sigma^2_A
&\approx&\sum_{S^x_N,S'^x_N}P^{\lambda=\infty}_{\frac{N}{2},S^x_N}P^{\lambda=\infty}_{\frac{N}{2},S'^x_N}\sum_{m\neq m'}\frac{\Big|\langle \Phi^{S^x_N}_m|A|\Phi^{S'^x_N}_{m'}\rangle\Big|^2}{{N \choose N/2+S^x_N}{N \choose N/2+S'^x_N}},\ \ 
\eea
where $P^{\lambda=\infty}_{\frac{N}{2},S^x_N}\equiv\big|\langle\psi(0)|\Phi^{\lambda=\infty}_{\frac{N}{2},S^x_N}\rangle\big|^2$ is the probability of initial pre-quenched state $|\psi(0)\rangle=|\!\downarrow,\downarrow,\cdots,\downarrow\rangle_z$ on the LMG eigenstate $|\Phi^{\lambda=\infty}_{\frac{N}{2},S^x_N}\rangle$. The above expression is the Eq.~(5) in the main text.

By further defining the quantity of
\bea\label{overlineA2}
\overline{A^2}_{S^x_N,S'^x_N}&\equiv&\sum_{m\neq m'}\frac{\Big|\langle \Phi^{S^x_N}_m|A|\Phi^{S'^x_N}_{m'}\rangle\Big|^2}{{N \choose N/2+S^x_N}{N \choose N/2+S'^x_N}}\nonumber\\
&=&\sum_{m, m'}\frac{\Big|\langle \Phi^{S^x_N}_m|A|\Phi^{S'^x_N}_{m'}\rangle-\langle \Phi^{S^x_N}_m|A|\Phi^{S'^x_N}_{m'}\rangle\delta_{mm'}\Big|^2}{{N \choose N/2+S^x_N}{N \choose N/2+S'^x_N}},\nonumber\\
\eea
we obtain the compact form of Eq.~(\ref{sigma2App})
\bea\label{sigmaA2}
\sigma^2_A
&\approx&\sum_{S^x_N,S'^x_N}P^{\lambda=\infty}_{\frac{N}{2},S^x_N}P^{\lambda=\infty}_{\frac{N}{2},S'^x_N}\overline{A^2}_{S^x_N,S'^x_N}.
\eea
In order to further simply Eq.~(\ref{sigmaA2}), one needs to calculate the matrix element of $\langle \Phi^{S^x_N}_m|A|\Phi^{S'^x_N}_{m'}\rangle$. However, the TFI Hamiltonian is non-integrable for $\alpha\neq 0$. Although it is impossible to obtain an exact analytical expression for a generic non-integrable Hamiltonian, we can calculate it approximately using the so-called eigenstate thermalisation hypothesis (ETH)\cite{Deutsch1991,Srednicki1994,Srednicki1999,Beugeling2015,Takashi2018}.
Following ETH, we write the matrix element of TFI Hamiltonian ($\alpha\neq0$)
\bea\label{ETH}
\langle \Phi^{S^x_N}_m|A|\Phi^{S'^x_N}_{m'}\rangle\approx \overline{A}\delta_{mm'}+\sqrt{\frac{\overline{A^2}}{D}}R_{mm'}
\eea
where $\overline{A}$ and $\overline{A^2}$ are smooth functions of energy levels, $R_{mm'}$ is a random variable with zero mean and unit variance, $D$ is the many-body Hilbert space dimension. In Fig.~\ref{fig_LevelStructure}(d), we plot the matrix elements of observable $A=N^{-1}\sum_j\sigma^z_j=2N^{-1}S^z_N$ in the basis of TFI Hamiltonian eigenstates $|\Phi^{S^x_N}_m\rangle$.
Since the observable $A=2N^{-1}S^z_N$ flips the total spin along $x-$direction, only the nearest-neighboring $S^{x}_N$ subspaces (i.e., $S'^{x}_N=S^x_N\pm 1$) are coupled via operator $A$. As a result, the whole matrix is divided into several blocks representing the nonzero coupling between symmetry sectors of $S^x_N$. The matrix elements of these blocks can be described by ETH (\ref{ETH}), but the variance $\overline{A^2}$ depends on the block and the relevant Hilbert dimension can be set as $D={N \choose N/2+S^x_N}+{N \choose N/2+S'^x_N}$.
We see that the quantity $\overline{A^2}_{S^x_N,S'^x_N}$ given by Eq.~(\ref{overlineA2}) is just the variance of the random part in the matrix element Eq.~(\ref{ETH}), which can be calculated 
\bea
 \overline{A^2}_{S^x_N,S'^x_N}=\frac{\overline{A^2}}{{N \choose N/2+S^x_N}+{N \choose N/2+S'^x_N}}.
\eea
Therefore, we have the variance of temporal fluctuations from Eq.~(\ref{sigmaA2})
\bea\label{AnalyticalFormula}
\sigma^2_A
&\approx&\sum_{S^x_N, S'^x_N}\frac{P^{\lambda=\infty}_{\frac{N}{2},S^x_N}P^{\lambda=\infty}_{\frac{N}{2},S'^x_N}}{{N \choose N/2+S^x_N}+{N \choose N/2+S'^x_N}}\overline{A^2}.
\eea
Notice that the pre-factor $\overline{A^2}$ in the above expression is still not determined yet.

To estimate the pre-factor $\overline{A^2}$ in Eq.~(\ref{AnalyticalFormula}), we first calculate $\sigma^2_A$ near the LMG model ( i.e., $\alpha\rightarrow0$ but $\alpha\neq 0$). In this limit, we can assume all the energy levels are lifted but the eigenstates are still the ones of LMG model \cite{Florian2017}. In this case, since the initial state is prepared with all the spins down along $z$-direction, we only need to consider the subspace with the total spin quantum number $S_N=N/2$ as shown by the first column in Fig.~\ref{fig_LevelStructure}(a). Therefore, we have the  fluctuation directly from Eq.~(\ref{sigma2A2})
\bea\label{AnalyticalFormula2}
\sigma^2_A
&=&\sum_{S^x_N, S'^x_N}{P^{\lambda=\infty}_{\frac{N}{2},S^x_N}P^{\lambda=\infty}_{\frac{N}{2},S'^x_N}}\Big|\langle \Phi^{\lambda=\infty}_{\frac{N}{2},S^x_N}|A|\Phi^{\lambda=\infty}_{\frac{N}{2},S'^{x}_N}\rangle\Big|^2.
\eea
The above expression (\ref{AnalyticalFormula2}) can be considered as the special case of the expression (\ref{AnalyticalFormula}). 
We emphasize that Eq.~(\ref{AnalyticalFormula2}) is NOT valid at at the integrable point $\alpha=0$ of LMG model since it is directly deduced from Eq.~(\ref{sigma2Asm}) (i.e., Eq.(3) in the main text) based on the assumption that all the energy levels are lifted. For the integrable LMG model which has exponentially many degeneracies with the number of spins, the temporal fluctuation is given by Eq.~(2) in the main text.
We need to modify Eq.~(\ref{AnalyticalFormula}) before connecting it to Eq.~(\ref{AnalyticalFormula2}). As discussed above, the LMG eigenstate $|\Phi^{\lambda=\infty}_{\frac{N}{2},S^x_N}\rangle$ is only energy level in the $S^x_N$ subspace coupled to the pre-quenched state $|\psi(0)\rangle=|\!\downarrow,\downarrow,\cdots,\downarrow\rangle_z$. In other words, all the relevant blocks of observable $A$ matrix to calculate $\sigma_A$, as shown in Fig.~\ref{fig_LevelStructure}(d), are reduced into $2\times 2$ matrices. As a result, the denominator in Eq.~(\ref{AnalyticalFormula2}), i.e., the Hilbert space dimension of $2\times 2$ blocks, should be replaced by $D=2$. By comparing the modified Eq.~(\ref{AnalyticalFormula}) to the Eq.~(\ref{AnalyticalFormula2}) at integrable point, we extract the pre-factor 
$$
\overline{A^2}=2\Big|\langle \Phi^{\lambda=\infty}_{\frac{N}{2},S^x_N}|A|\Phi^{\lambda=\infty}_{\frac{N}{2},S'^{x}_N}\rangle\Big|^2.
$$
Finally, we obtain the formula (6) in the main text, i.e.,
\bea\label{sigma2}
&&\sigma^2_A\approx 2\sum_{S^x_N,S'^x_N}\frac{P^{\lambda=\infty}_{\frac{N}{2},S^x_N}P^{\lambda=\infty}_{\frac{N}{2},S'^x_N}}{{N \choose N/2+S^x_N}+{N \choose N/2+S'^x_N}}\Big|A^{\lambda=\infty}_{S^x_NS'^x_N}\Big|^2
\eea
with the matrix element defined by $$ A^{\lambda=\infty}_{S^x_NS'^x_N}\equiv\langle \Phi^{\lambda=\infty}_{\frac{N}{2},S^x_N}|A|\Phi^{\lambda=\infty}_{\frac{N}{2},S'^{x}_N}\rangle.$$
As mentioned in the main text, we expect the analytical formula (\ref{sigma2}) still works for a finite value of $\lambda$, but the eigenstate $|\Phi^{\lambda}_{\frac{N}{2},S^x_N}\rangle$ should be referred to the energy level adiabatically connecting to $|\Phi^{\lambda=\infty}_{\frac{N}{2},S^x_N}\rangle$.

\begin{figure}
  \includegraphics[scale=0.3]{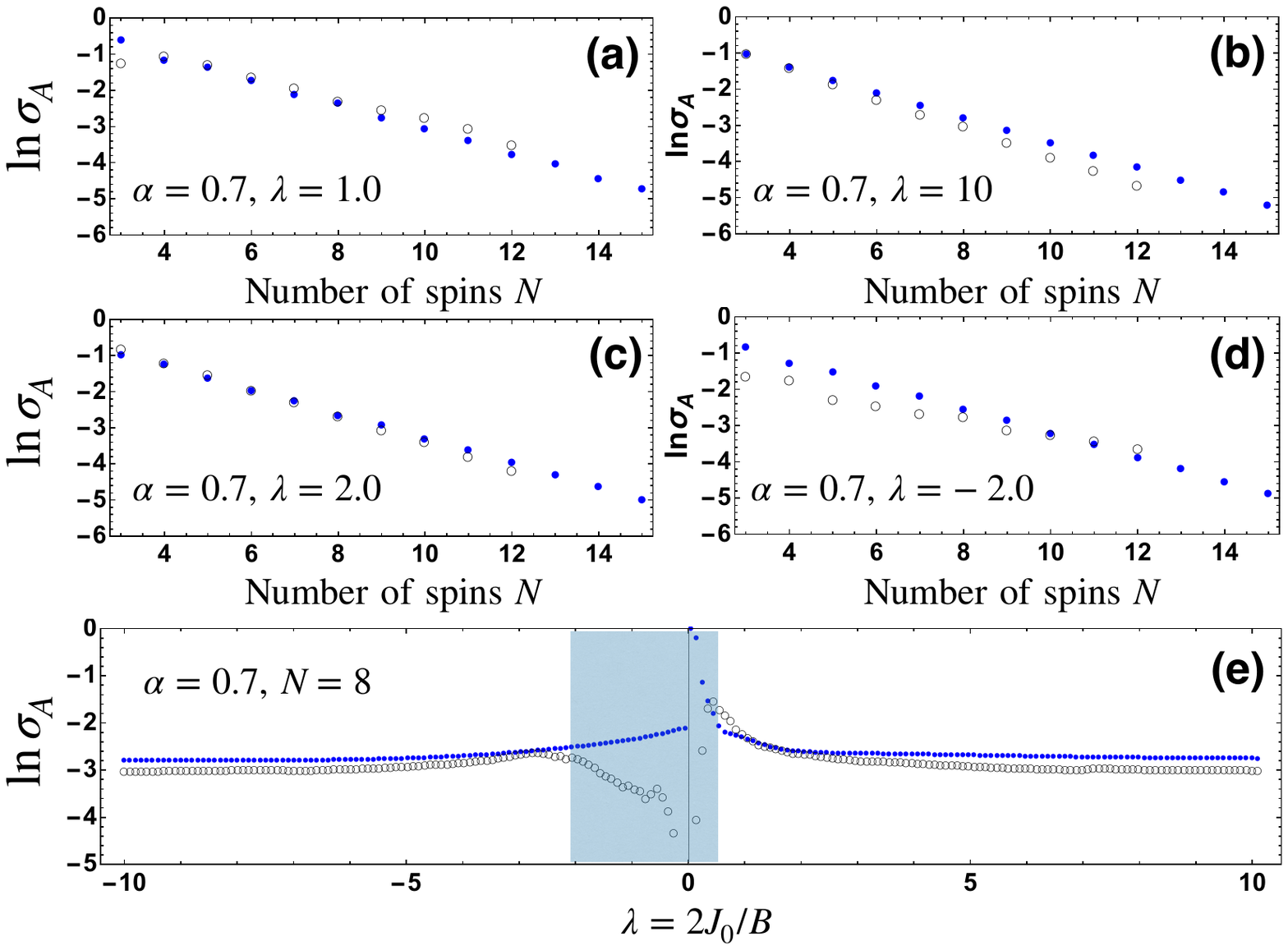}
  \caption{ Compare the temporal fluctuations $\ln\sigma_A$ calculated from exact diagonalization up to $N=12$ spins (black empty circles) and analytical formula Eq.~(\ref{sigma2}) up to $N=15$ spins (blue solid dots).  
 Figs.~(a-d) show $\ln\sigma_A$ as functions of system size $N$ while Fig.~(e)   shows $\ln\sigma_A$ as a function of $\lambda$. }
  \label{fig_FormulaCheck}
\end{figure}

We emphasize that several physical assumptions are made in the derivation of 
Eq.~(\ref{sigma2}) including the Berry's conjecture, ETH and the connection to LMG model. The validity of Eq.~(\ref{sigma2}) is verified numerically as shown by Fig.~\ref{fig_FormulaCheck}. In Figs.~(a)-(d), we compare the temporal fluctuations $\ln\sigma_A$ as function of $N$ calculated from exact diagonalization (black empty circles) and analytical formula (\ref{sigma2}) (blue solid dots) for different parameter settings. They agree with each other very well. 
As discussed in the main text, Eq.~(\ref{sigma2}) is valid in the parameter regime $\alpha\ll\alpha^*=\ln(2|\lambda|)/\ln2.$ Thus, for a fixed $\alpha>0$, it needs that $|\lambda|\gg 2^{\alpha-1}$.  
In Fig.~\ref{fig_FormulaCheck}(e), we show that Eq.~(\ref{sigma2}) fails when the parameter $\lambda$ is in the regime $-2.0\lesssim\lambda\lesssim 0.5$ for $\alpha=0.7$ as indicated by the dark region. 
 For the parameter $|\lambda|\leq0.5$, the critical value of power-law exponent $\alpha^*\leq 0$; thus no value of $\alpha >0$
satisfies Eq.~(\ref{sigma2}).

\section{{II. Trapped-Ion experimental system}}\label{TechnicalDetails}
The ions are trapped in a macroscopic linear Paul trap with transverse center-of-mass (COM) trap frequency $\nu_{COM}=4.4$ MHz. The trap is housed in a cryogenic vacuum chamber in order to reduce the background vacuum pressure \cite{CryoTechnical}.
To conduct the experiment, the spins are initialized into the $|\!\downarrow\downarrow...\downarrow\rangle_z$ state by optical pumping with resonant 369.5 nm light. Coherent spin rotations and spin-spin interactions are performed using 355 nm counter-propagating Raman beams.

In order to generate the Hamiltonian (1) in the main text, we use the M\o{}lmer-S\o{}rensen (MS) protocol \cite{Molmer} by applying a  bichromatic Raman beatnote at frequencies $\omega_0\pm\mu$, where $\omega_0/2\pi= 12.643$ GHz is the qubit frequency. The bichromatic beat-notes off-resonantly excite the transverse modes of motion generating the Ising Hamiltonian \cite{Kim2009}.
\begin{equation}
\label{eq_H}
H = \sum_{i<j}J_{ij}\sigma^x_i \sigma^x_j, \,\,\, J_{ij} = \Omega^2\omega_{R} \sum_{m}\frac{b_{im}b_{jm}}{\mu^2-\omega_m^2}.
\end{equation}
Here $\Omega$ is the global carrier Rabi frequency coupling the electronic states $\ket{\downarrow}_z$ and $\ket{\uparrow}_z$, $\omega_R= \hbar \Delta  k^2/(2M)$ is the recoil frequency, $b_{im}$ is the normal mode transformation matrix element of the $i$-th ion with the $m$th normal mode, $\Delta k$ is the difference wave-vector between the two Raman beat-notes, $M$ is the mass of a single ion, and $\omega_m$ is the frequency of the $m$-th normal mode.
Equation (\ref{eq_H}) shows that the collective modes of vibration of the ion chain mediate the long-range spin-spin coupling.
The transverse field, along $\hat z$, is created by $\pm \mu\rightarrow\pm\mu+B$ to the red and blue Raman beat-notes, generating an effective magnetic field of strength $B/2$ \cite{AaronThesis}.

At the end of each experimental sequence, we measure each spin's magnetization with spin-dependent fluorescence using Andor iXon Ultra 897 EMCCD camera. A 369.5 nm laser resonant with the $^2$S$_{1/2}\ket{F=1}\leftrightarrow {^2}$P$_{1/2}\ket{F=0}$ transition causes photons to scatter off each ion if the qubit is projected to the $\ket{\uparrow}_z$ `bright' state. Conversely, ions projected to $\ket{\downarrow}_z$ `dark' state scatter negligible number of photons as the laser is detuned from the resonance by the $^2$S$_{1/2}$ hyperfine splitting \cite{Olmschenk2007}.


\begin{table}
\begin{tabular}{ |p{0.6cm}|p{1.3cm}|p{2.7cm}|p{3.2cm}|  }
 \hline
 \multicolumn{4}{|c|}{Experimental Details} \\
 \hline
 Ions & Mean $\alpha$ & Total exp. points &Range of $J_0/2\pi$ (\si{\kilo\hertz}) \\
 \hline
 3 &  0.725815  & 1000 & 0.53 - 0.541\\
 4 &  0.709570  & 1000 & 0.50 - 0.60\\
 5 &  0.692255  & 1000 & 0.45 - 0.55\\
 6 &  0.678499  & 1200 & 0.41 - 0.53\\
 7 &  0.664681  & 1200 & 0.39 - 0.52\\
 8 &  0.648291  & 1400 & 0.38 - 0.48\\
 \hline
\end{tabular}
\caption{Experimental values used for $N=3$ to $8$ ions.\label{Table1}}
\end{table}



We work in the far-detuned regime ($\mu-\omega_{COM}\gg\eta \Omega_{COM}$, where $\eta=\sqrt{\omega_R/\omega_{COM}}$ is the Lamb-Dicke factor), in order to reduce the residual spin-motion entanglement, caused by off-resonant excitation of the ion chain's motional modes \cite{Wang2012}. Residual spin-motion entanglement results in bit-flip errors on the spin as motional degrees of freedom are traced out at the end of the experiment. The probability of this error to occur on the $i$th ion is proportional to $p_i \approx \sum^N_{m=1} (\eta_{im} \Omega/ \delta_m)^2$, where $\eta_{im} = b_{im}\sqrt{\omega_R/\omega_{m}}$ and  $\delta_{m} = \mu-\omega_m$ is the beatnote detuning from the $m$th motional mode \cite{YukaiThesis}. To minimize this error, we choose $\delta_{COM}$ such that $(\eta_{COM}\Omega / \delta_{COM})^2 \lesssim 1/9$. 

As we are working in the far-detuned regime, the spin-spin interaction is reduced, making the system susceptible to slow noise, both in $J_0$ and in $B$. Therefore, in the course of data collection, we routinely balance the differential light shift generated by the red and blue Raman beatnotes, that creates an offset in the effective magnetic field $B$. While the light shift is fairly stable over the course of one experimental scan (which takes $\sim 2$ minutes), the net light shift can change between different scans, mainly because of noise in beam pointing, intensity and trap frequency. This drift is detected with a Ramsey experiment and is calibrated out, resulting in a standard deviation $\sigma_{B}$ of about $2\pi\times\SI{0.1}{ \kilo\hertz}$. The nearest neighbour spin-spin interaction $J_0$ was also measured before and after taking a set of data and will drift by around 5\% peak-to-peak. 




After the quench with Hamiltonian (1) in the main text, the spins are allowed to evolve and are measured at 50 time-steps between (0-2.0) ms. The magnetization of the spins is measured in the $\hat z$ basis. During a given scan, 200 experiments per time-step are taken. Each experimental scan is repeated 5-7 times (Table \ref{Table1}) in order to have a large sample which will suppress the measurement noise. This allows for the detection of persistent fluctuations.

Table \ref{Table1} summarizes the experimental values used. The number of ions used were 3 to 8 ions and the number of total experiments taken increased with ion number since the fluctuation signal decreases as ion number increases.

For each ion, and for each time step, all of the data points are averaged 1000-1400 times depending on $N$ (Table \ref{Table1}).
Then, all of the ions in a given time step are averaged.
This results in data that are plotted as average magnetization as a function of time as shown in Fig.~2 in the main text. The standard deviation of the last 48 (out of 50 total) steps is taken to characterize the temporal fluctuations.

The standard deviation of the average magnetization can be plotted as a function of $\lambda = 2J_0/B$, where $B$ takes on the values of $\pm2\pi\times(0.5, 0.75, 1.0, 1.5, 2.0)$ \si{\kilo\hertz} as shown by Fig.~3 in the main text. Data at some other $B$ values are used to better resolve features of the curve.
The standard deviation of temporal fluctuations was calculated from the last 48 time steps (from Fig.~2 in the main text).
The standard deviation and the error bars on those values (Fig.~3 in the main text) were calculated as described in the next section.
In Fig.~4 in the main text, the data points with fixed values of $\lambda=2J_0/B$ are plotted for different numbers of ions in the experiment and compared to the numerical results.

\section{III. Measurement error bars}

\subsection{A. Error bars in Fig.~2 of main text} 

The observable we measured is the average magnetization of $N$ spins given by
\bea
A(t_i)=\frac 1 N\sum^N_{j=1} \sigma^z_j(t_i).
\eea
Here, $\sigma^z_j(t_i)$ represents the magnetization of spin $j$ at time step $t_i$, which is obtained by performing $M$ measurements. Each measured value of $\sigma^z_{j,k}(t_i)$ with measurement index $(k=1,2,\cdots,M)$ is a random binary number, i.e., $\sigma^z_{j,k}(t_i)$ is $+1$ for spin up and $-1$ for  spin down. The {\it sample mean} is given by
\bea\label{samplemean}
\overline{\sigma^z_{j,M}(t_i)}\equiv\frac 1 M\sum^M_{k=1} \sigma^z_{j,k}(t_i).
\eea
In the limit of $M\rightarrow\infty$, the sample mean $\overline{\sigma^z_{j,M}(t_i)}$ converges to the fixed {\it population mean} value $\sigma^z_j(t_i)$. 
The {\it sample variance} of measured values is given by
\bea\label{samplevariance}
\overline{\sigma^{z}_{j,M}(t_i)^2}\equiv\frac 1 M\sum^M_{k=1} \Big(\sigma^z_{j,k}(t_i)-\overline{\sigma^z_{j,M}(t_i)}\Big)^2.
\eea
In the limit of $M\rightarrow\infty$, the sample variance $\overline{\sigma^{z}_{j,M}(t_i)^2}$ converges to the {\it population variance} $\mathrm{Var}[\sigma^{z}_{j}(t_i)]$. 

However, in the experiments, the number of measurements $M$ is finite. As a result, both $\overline{\sigma^z_j(t_i)}$ and $\overline{\sigma^{z}_{j,M}(t_i)^2}$ are random variables. The expected value of $\overline{\sigma^{z}_{j,M}(t_i)^2}$ can be calculated from Eqs.~(\ref{samplemean}) and (\ref{samplevariance}), i.e.,
\bea
E\Big[\overline{\sigma^{z}_{j,M}(t_i)^2}\Big]=\frac{M-1}{M}\mathrm{Var}[\sigma^{z}_{j}(t_i)].
\eea
 Hence, $\overline{\sigma^{z}_{j,M}(t_i)^2}$ gives an estimate of the population variance that is biased by a factor of $(M-1)/M$. For this reason, $\overline{\sigma^{z}_{j,M}(t_i)^2}$ is referred to as the {\it biased sample variance}. The {\it unbiased sample variance} is defined by 
\bea
\overline{v^{z}_{j,M}(t_i)^2}&\equiv&\frac{M}{M-1}\overline{\sigma^{z}_{j,M}(t_i)^2}
\nonumber\\
&=&\frac{1}{M-1}\sum^M_{k=1} \Big(\sigma^z_{j,k}(t_i)-\overline{\sigma^z_{j,M}(t_i)}\Big)^2.
\eea

The sample mean $\overline{\sigma^z_{j,M}(t_i)}$ is also a random variable due to finite sampling. By repeated sampling and recording of the means obtained, a sampling distribution of different means is generated. This distribution has its own mean and variance. The standard deviation of this sampling distribution is called the {\it standard error of the mean} (SEM), which is given by
\begin{equation}
\Sigma_{\overline{\sigma^z_{j}(t_i)}}=\sqrt{\frac{\mathrm{Var}[\sigma^{z}_{j}(t_i)]}{M}}\approx\sqrt{\frac{\overline{v^{z}_{j,M}(t_i)^2}}{M}}
\label{SEM}
\end{equation}
Since the population variance $\mathrm{Var}[\sigma^{z}_{j}(t_i)]$ is seldom known, we have replaced it by the unbiased sample variance $\overline{v^{z}_{j,M}(t_i)^2}$. 

In the end, we calculate and plot the sample mean of $A(t_i)$ by finite measurements, i.e.,
\bea
\overline{A_{M}(t_i)}=\frac{1}{M}\sum_{k=1}^MA_k(t_i)=\frac 1 N\sum^N_{j=1} \overline{\sigma^z_{j,M}(t_i)}.
\eea
Here, $A_k(t_i)$ is the value of $k$-th measurement at time step $t_i$. 
The average magnetization per time step, $\overline{A_{M}(t_i)}$, was calculated by averaging the magnetization per ion over the number of ions in the experiment for each time step where each spin is weighted evenly. Using the number of experimental shots given in Table~\ref{Table1} and the average magnetization of a given ion, the standard deviation of the corresponding Binomial distribution was obtained for each ion at a given time step, i.e., $\Sigma_{\overline{\sigma^z_j(t_i)}}$ for the $j-$th ions.
The error in $\overline{A_{M}(t_i)}$ is produced by taking the values $\Sigma_{\overline{\sigma^z_{j}(t_i)}}$ and adding them in quadrature with even weight using the equation
\begin{equation}\label{SigmaAM}
\Sigma_{\overline{A_{M}(t_i)}} = \sqrt{
\sum_{j=1}^{N}  \left( \frac{1}{N}\Sigma_{\overline{\sigma^z_{j}(t_i)}} \right) ^2}.
\end{equation}
The quantity $\Sigma_{\overline{A_M}}(t_i)$ is plotted as the error bar for each data point shown in Fig.2 in the main text.

\subsection{B. Error bars in Fig.~3 of main text}

The temporal fluctuations are the values of $\overline{A_{M}(t_i)}$ for different time steps. For given time interval $t_i\in [t_0,t_0+T]$, the mean of temporal fluctuations is defined by
\begin{equation}
\langle A\rangle\equiv\frac{1}{T}\int_{t_0}^{t_0+T}\overline{A_{M}(t_i)}dt.
\end{equation}
The variance of temporal fluctuations is defined by
\begin{equation}
\sigma^2_A\equiv\frac{1}{T}\int_{t_0}^{t_0+T}\Big(\overline{A_{M}(t_i)}-\langle A\rangle\Big)^2dt.
\end{equation}
In the experiment, we only measure $A$ at $n$ discrete time steps and use them to calculate the mean of temporal fluctuations
\begin{equation}
\langle A_n\rangle=\frac 1 n\sum^n_{i=1}\overline{A_{M}(t_i)}
\end{equation}
and the standard deviation of temporal fluctuations
%
\begin{equation}\label{sigmaAn}
\sigma_{A,n}=\sqrt{\frac{1}{n}\sum^n_{i=1} \Big(\overline{A_{M}(t_i)}-\langle A_n\rangle\Big)^2}.
\end{equation}
To obtain the SEM of the quantity Eq.~(\ref{sigmaAn}), we take the derivative of Eq.~(\ref{sigmaAn}) with respect to $\overline{A_{M}(t_i)}$, i.e., 
\bea
D_{\sigma_{A,n}}(t_i)&=&\frac{1}{2}[\sigma_{A,n}^2]^{-1/2}\frac{2}{n}[\overline{A_{M}(t_i)}-\langle A_n\rangle] \nonumber\\
&=&\frac{\overline{A_{M}(t_i)}-\langle A_n\rangle}{n\sigma_{A,n}}.
\eea
The final error value used to plot Fig.~3 in the main text is then obtained by adding the values of $D_{\sigma_{A,n}}(t_i)$ multiplied by $\Sigma_{\overline{A_{M}}}(t_i)$ in quadrature (over the last 48 data points of Fig.~2 in the main text), i.e., 
\begin{equation}
\Sigma_{\sigma_{A,n}} = \sqrt{\sum_{t_i=3}^{t_i=50}\Big[D_{\sigma_{A,n}}(t_i)\Sigma_{\overline{A_{M}(t_i)}}\Big]^2}.
\end{equation}

\begin{figure}
\includegraphics[width=\linewidth]{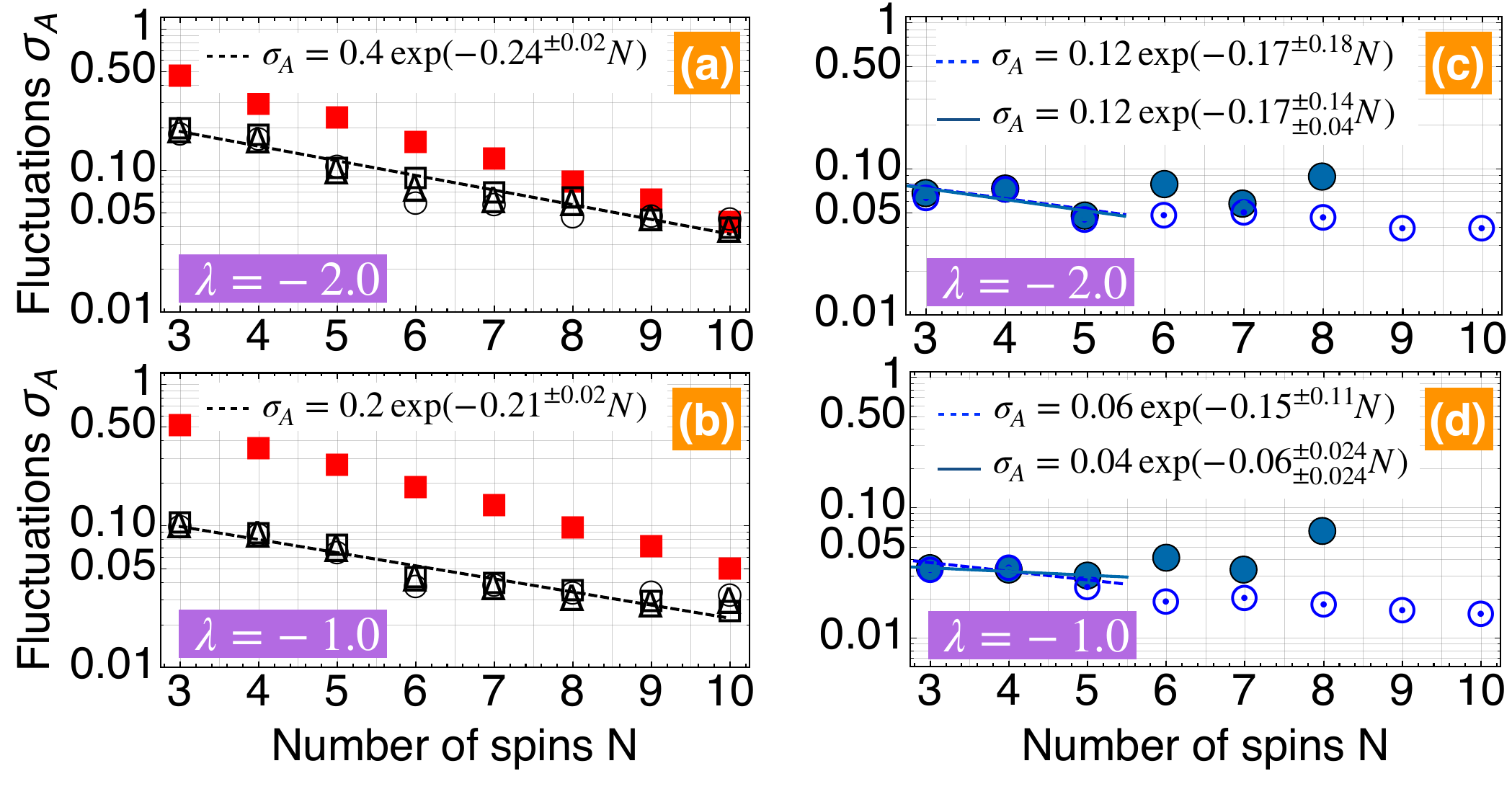}
\caption{\label{fig_lnsigmavsN1}
 {Logplot of temporal fluctuation $\sigma_A$ versus number of spins $N$ for different values of $\lambda$:} (a-b) Analytical results (red squares) vs numerical results with different time windows $Jt\in [0,2\pi]$ (circles), $Jt\in [0,5\pi]$ (triangles) and $Jt\in [0,\infty]$ (empty squares) for $\alpha=0.7$.
(c-d) Experimental results (blue dots) vs
numerical results (circled dots) taken from Fig.~3 in the main text. The black dashed lines in Figs.~(a-b) are the fits to the numerical infinite-time-window averaging data, while the solid blue lines and the dashed blue lines in Figs.~(c-d) are the fits to the experimental data and numerical data respectively.}
\end{figure}

\begin{figure}
\includegraphics[width=\linewidth]{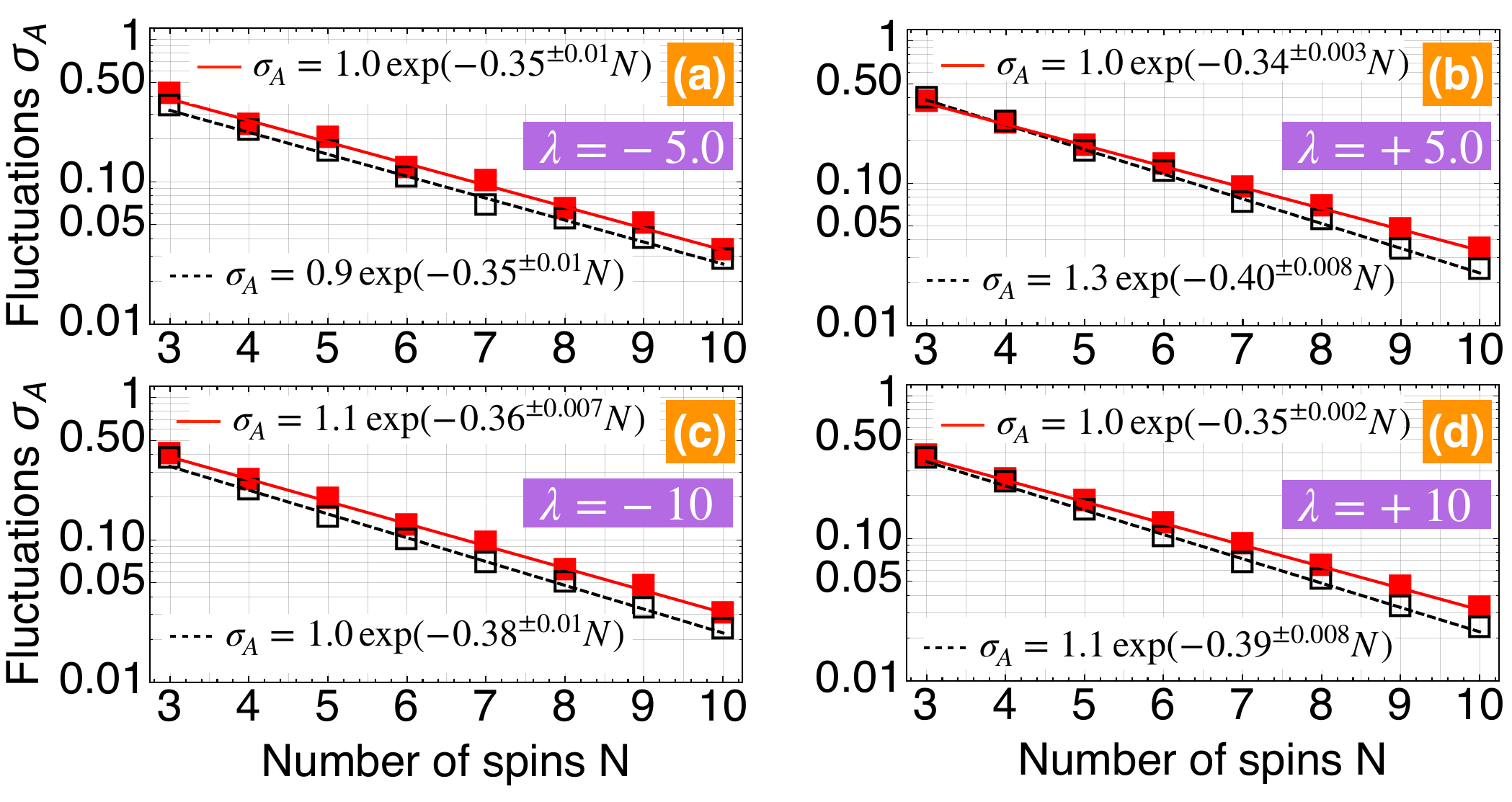}
\caption{\label{fig_lnsigmavsN2}
 {Logplot of temporal fluctuation $\sigma_A$ versus number of spins $N$ for large values of $\lambda=\pm 5.0$ (a-b) and $\lambda=\pm 10.0$ (c-d):} analytical results (red squares) vs numerical results with infinite time window averaging (empty squares) for $\alpha=0.7$. The solid red lines and the dashed black lines are the fits to the analytical results and the numerical infinite-time-window averaging results respectively.}
\end{figure}

\section{{IV. System size scaling}}\label{Sizescaling}

\subsection{A. Extracting scaling exponent}

In this section, we discuss how the system size scaling exponents shown in Fig.~4 in the main text are extracted from the numerical and experimental data. We clarify the uncertainties of scaling exponent and show plots for more values of $\lambda$.

First, we introduce the linear least squares fitting technique. Given a set of data points from numerical simulations or experimental measurements, $$\{(x_i,y_i)\mid i=1,2,\cdots,n\},$$ we search for a linear function $$f(x,a,\kappa)=a+b x$$ to fit the data points. To get the best fit line, we define the error function
\bea
R^2=\sum_{i=1}^{n}[y_i-f(x_i,a,b)]^2.
\eea
The minimum of $R^2$ is determined by $\partial R^2/\partial a=0$ and $\partial R^2/\partial b=0$. The final results are
\bea
b=ss_{xy}/ss_{xx}, \ \ \ \ \ a=\bar{y}-b\bar{x},
\eea
where $\bar{y}=\sum_{i=1}^ny_i/n$, $\bar{x}=\sum_{i=1}^nx_i/n$, and 
\bea
ss_{xy}&\equiv& \sum_{i=1}^n(x_i-\bar{x})(y_i-\bar{y})\nonumber\\
ss_{xx}&\equiv& \sum_{i=1}^n(x_i-\bar{x})^2\nonumber\\
ss_{yy}&\equiv& \sum_{i=1}^n(y_i-\bar{y})^2.
\eea
Since there are residual errors between the actual data point $y_i$ and the best-fitted point $f(x_i,a,b)$, i.e., $e_i=y_i-(a+b x_i)$, one can define $s$ as an estimator for the errors
\bea
s=\sqrt{\sum_{i=1}^n\frac{e_i^2}{n-2}}=\sqrt{\frac{ss_{yy}-\frac{ss^2_{xy}}{ss_{xx}}}{n-2}}.
\eea
The {\it standard error} of $b$ is given by \cite{LSF}
\bea\label{SEb}
\mathrm{SE}(b)=s/\sqrt{ss_{xx}}
\eea
which is {\it the uncertainty of $b$ from the least square fitting}. 
If each data point $y_i$ further has statistical error with variance $\mathrm{Var}[y_i]$, there is additional uncertainty of $b$
\bea
\Sigma(b)=\sqrt{\sum_{i=1}^n\Big(\frac{\partial b}{\partial y_i}\Big)^2\mathrm{Var}[y_i]}
\eea
which is {\it the uncertainty of $b$ from the statistical errors}.

In Fig.~4 in the main text, we have shown the system size scaling for positive $\lambda>0$ values. Here, in Fig.~\ref{fig_lnsigmavsN1} and Fig.~\ref{fig_lnsigmavsN2}, we show the plots for negative $\lambda<0$ values.
We choose the number of spins $N$ as $x_i$ data and the logarithm of temporal fluctuations $\ln\sigma_A$ as $y_i$ data.
We show the fits to both the experimental data and numerical results. We indicate the uncertainties of system size scaling exponent ($\kappa=-b$) from the least square fitting (superscript) and the statistical errors (subscript) in the experiments.  
%

%

In Figs.~\ref{fig_lnsigmavsN1}(a)-(b), we compare the analytical results from Eq.~(6) in the main text with the numerical results with infinite-time-window averaging data. In Figs.~\ref{fig_lnsigmavsN1}(c)-(d), we compare the experimental data with the numerical results taken for $\lambda=-2.0$ and $\lambda=-1.0$. 

In Figs.~\ref{fig_lnsigmavsN2}, we extract the system size scaling exponents from both analytical results and numerical results for $\lambda=\pm 5.0$ and $\lambda=\pm 10.0$. Our analytical results agree with numerical simulations, and the extracted system size exponents are close to the theoretical prediction $\kappa=\ln\sqrt{2}$ for $N\gg 1$. In fact, our analytical expression is valid in the regime of $|\lambda\gg 2^{\alpha-1}|$ as discussed in Section \ref{TF} and shown in Fig.~\ref{fig_FormulaCheck}(e). 

\subsection{B. Student's t-test}

In this section, we perform the Student’s t-test to estimate the likelihood of measuring the observed slopes under the null hypothesis of single-particle dephasing. 
For the single-particle dephasing, the temporal fluctuations follow the power-law $\sigma_A\propto 1/\sqrt{N}$ \cite{Florian2017}, an exponential fit of data generated from single-particle dephasing for $N =3-5$
would lead to an exponent $\kappa_0\sim 0.13$. In the fits to the experimental data for some $\lambda$ values, we obtain larger scaling exponents, e.g., $\kappa=0.25$ for $\lambda=1.7$, $\kappa=0.17$ for $\lambda=1.3$ and $\lambda=-2.0$. However, due to the small number of samples, i.e., $n=3$ ($\lambda=1.7, -2.0$) or $n=4$ ($\lambda=1.3$), we need to consider the standard error $\Sigma(\kappa)$, given by Eq.~(\ref{SEb}), of the exponent $\kappa$ extracted from the data. We take the single-particle-dephasing exponent $\kappa_0=0.13$ as the {\it null hypothesis} and calculate the statistical quantity
\bea
t_\kappa=\frac{\kappa-\kappa_0}{\mathrm{SE}(\kappa)},
\eea
which follows the Student's t-distribution with $n-2$ degrees of freedom. Then, we calculate the one-sided p-value 
\bea
p=\mathrm{Pr}(t>t_\kappa)=\int_{t_\kappa}^{+\infty}f(t)dt,
\eea
where $f(t)$ is the probability density function of Student's t-distribution given by
\bea
f(t)=\frac{\Gamma(\frac{\nu+1}{2})}{\sqrt{\nu\pi}\Gamma(\frac{\nu}{2})}\Big(1+\frac{t^2}{\nu}\Big)^{-\frac{\nu+1}{2}}
\eea
with $\nu$ the degrees of freedom ($\nu=n-2$ for the linear regression). The p-value is the probability of obtaining the measured scaling exponent $\kappa$ given that the null hypothesis of single-particle dephasing is true, purely due to random fluctuations. 

\begin{table}
\begin{tabular}{|p{1.0cm} |p{1.5cm}|p{1.5cm}|p{2cm}|p{1.5cm}|  }
 \hline
 \multicolumn{5}{|c|}{Student's t-test} \\
 \hline
 $\lambda$ & $\kappa$ & SE($\kappa$) & $t_\kappa$-statistic &  $p$-value \\
 \hline
 1.7 & 0.2475  & 0.0785 & 1.497 & 0.188\\
 1.3 &  0.1651  & 0.0252 & 1.393 & 0.149\\
 -2.0 &  0.1744  & 0.142 & 0.313 & 0.404\\
 \hline
\end{tabular}
\caption{Student's t-tests of system size scaling exponents for different $\lambda$ values. \label{Table2}}
\end{table}

In Table~\ref{Table2}, we display the p-values for different $\lambda$ values. The test for $\lambda=1.3$ has the smallest $p=0.149$, which means under the hypothetical assumption of single-particle dephasing the actually observed slope is unlikely to be encountered purely by random fluctuations. Considering the noises in $J_0$ and $B$, the single-particle-dephasing exponent would be smaller than $\kappa_0=0.13$. Thus, the statistical significance of the deviation from our actually observed slope increases when taking into account this effect.

It is also important to notice that the fits to the numerical results also have finite $p-$values. For example, as shown the Fig.~4(e) in the main text, the scaling exponent extracted from the numerical data for $\lambda=1.3$ is $\kappa=0.19$ with standard error $\mathrm{SE}(\kappa)=0.04$, which results in $t_\kappa=1.5$ and $p=0.136$. This means the theory itself does not predict a perfect linear fit if one takes into account experimental slow drifts. In fact, to fit the experimental data, we choose different $\alpha$ values for different ion numbers as displayed in Table~\ref{Table1}. The consistency of $p$-values between the experimental data test ($p=0.149$ for $\lambda=1.3$) and the numerical data test ($p=0.136$ for $\lambda=1.3$)  actually further confirms that our experiment supports our theory, while the theory predicts the extremely significant exponential system size scaling without noises as shown in Fig.4~(a) ($p=0.001$) and Fig.4~(b) ($p=0.0015$) in the main text. We note that for $\lambda=-2.0$ the $p$-value is higher with respect to the other two cases. We attribute this discrepancy to the asymmetry in the magnetization fluctuations as a function of $\lambda$. As explained in the next section, the amplitude of the fluctuations at negative $\lambda$ are lower than the respective positive $\lambda$ values. This means that the measurement is more sensitive to our noise baseline (see Fig. 3c) and the rejection of the null hypothesis has lower confidence. In the next section we will explain the origin of the asymmetry in the magnetization fluctuations as a function of $\lambda$.

\begin{figure*}
  \includegraphics[scale=0.35]{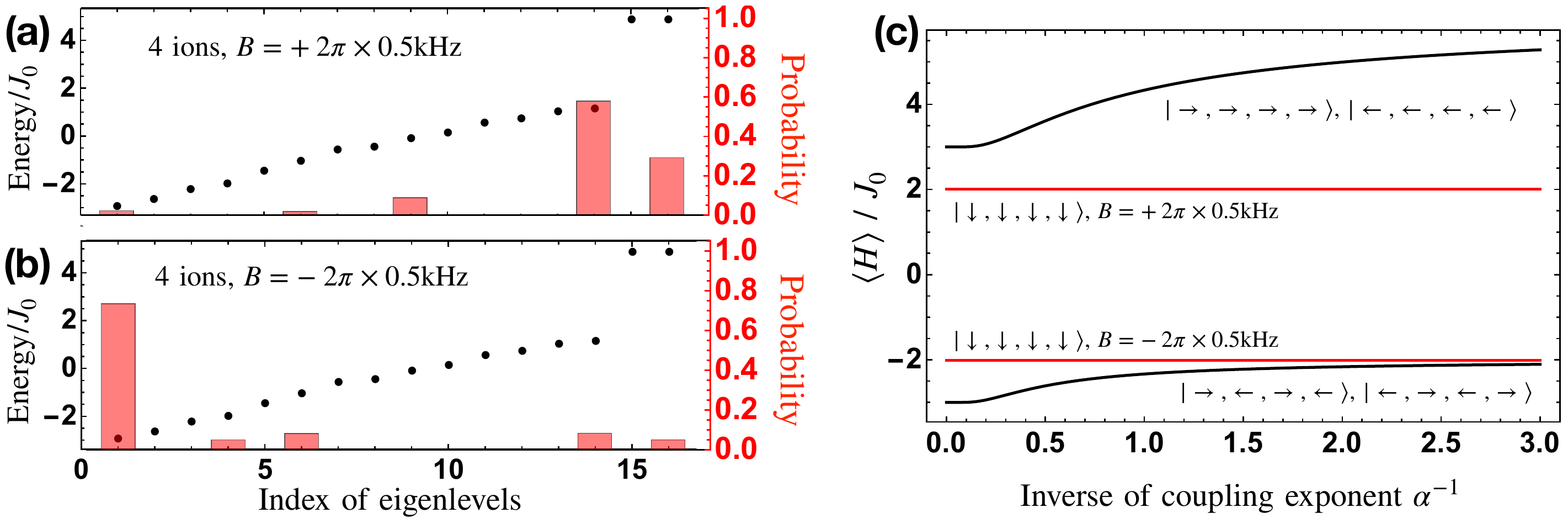}
  \caption{Energy level structure in the presence of long-range interaction. (a) and (b) Energy of eigenlevels with ascending order (black dots) and the probability distribution of initial state (red bars). Parameters: $J_0=2\pi\times\SI{0.5}{\kilo\hertz}$, $\alpha=0.73$. (c) The averaged energy of different spin states as a function of interaction range, i.e., the inverse of power-law coupling exponent $\alpha^{-1}$. }
  \label{fig_AsymmetricSpectrum}
\end{figure*}

 \begin{figure}
  \includegraphics[scale=0.24]{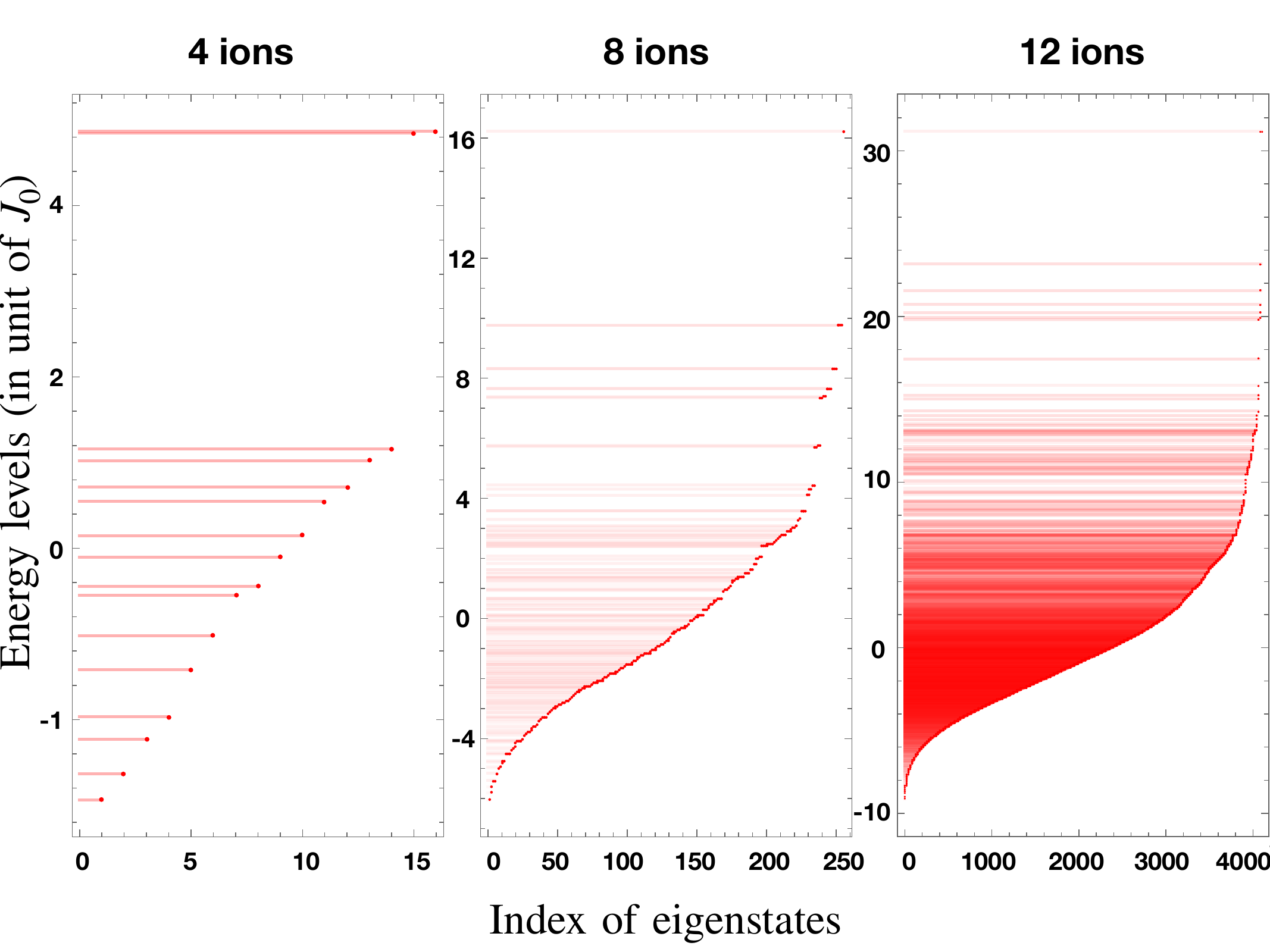}
  \caption{Energy level structure for different number of ions ($N=4, 8, 12$) with parameters $J_0=2\pi\times\SI{0.5}{\kilo\hertz}$, $\alpha=0.73$. }
  \label{fig_Gapsofspectrum}
\end{figure}

\section{{V. Asymmetric fluctuations}}\label{AsymmetricSpectrum}

We discuss in detail the asymmetric dynamical behaviours of the observable $\langle A(t) \rangle$  for the magnetic field with opposite signs. In Fig.~\ref{fig_AsymmetricSpectrum}(a) and (b), we plot the energy spectrum (black dots) with ascending order and the probability (red bars) of initial state (all spins down along $z-$direction) over the eigenstates for the magnetic field $B=+2\pi\times \SI{0.5}{\kilo\hertz}$ and $B=-2\pi\times \SI{0.5}{\kilo\hertz}$ respectively.

First, we see the energy spectra for opposite magnetic fields are identical. This is not difficult to understand since the Hamiltonian, i.e.,
\begin{equation}
H=J_0\sum_{i<j}\frac{1}{|i-j|^\alpha}\sigma_i^x\sigma_j^x - \frac{1}{2}B\sum_{i}\sigma_i^z,
\label{HamiltonianSM}
\end{equation}
is invariant by the transformation $B\rightarrow -B$ and $\vec{\sigma}_i\rightarrow -\vec{\sigma}_i$ for all the spins.

Second, the spectrum is asymmetric with respect to the zero value. At
high energies there is a significant energy gap, while at
low energies there is no obvious gap. This is due to the long range interaction. For the nearest-neighbor coupling ($\alpha=\infty$ or $\alpha^{-1}=0$), the spectrum is symmetric with respect to zero since the sign of Hamiltonian can be reversed by flipping all the spins along $z-$direction $\sigma^z_i\rightarrow -\sigma^z_i$ and changing the sign of all the neighboring coupling terms $\sigma_i^x\sigma_{i+1}^x$ by flipping every two spins along $x-$direction. In the case of positive coupling $J_0=2\pi\times\SI{0.5}{\kilo\hertz}$ and $B=\pm2\pi\times \SI{0.5}{\kilo\hertz}$, the lowest-energy state is the anti-ferromagnetic (AFM) state along $x-$direction which is superposition of AFM states $|\rightarrow,\leftarrow,\rightarrow,\leftarrow \rangle$ and $|\leftarrow,\rightarrow,\leftarrow,\rightarrow\rangle$, while the highest-energy state is the ferromagnetic (FM) state along $x-$direction which is superposition of FM states $|\rightarrow,\rightarrow,\rightarrow,\rightarrow \rangle$ and $|\leftarrow,\leftarrow,\leftarrow,\leftarrow\rangle$.

However, when the interaction range becomes longer by tuning $\alpha$ smaller, the averaged energy $\langle H \rangle$ over the FM states on the top of spectrum increases much faster than the AFM energy at the bottom of spectrum. This is shown by the two black curves in Fig.~\ref{fig_AsymmetricSpectrum}(c). For the AFM states, the different long-range coupling terms $\langle \sigma^x_i\sigma^x_j \rangle$($j>i$) have different signs and thus can cancel each other. However, for the FM states, all the long-range coupling terms have positive sign and thus increase the energy uniformly. As the longer interaction makes the spin flipping more difficult, a big energy gap appears the top of the energy spectrum.  

Third, the initial state with all spins down along $z-$direction (i.e., $|\downarrow,\downarrow,\downarrow,\downarrow\rangle$) stays on different sides of the spectrum depending on the sign of magnetic field $B$. For the positive magnetic field $B=+2\pi\times \SI{0.5}{\kilo\hertz}$, the averaged energy $\langle H \rangle$ is positive and the initial state is the superposition of several of the
highest exited states of the spectrum as shown in Fig.~\ref{fig_AsymmetricSpectrum}(a). The energy gap
leads to short-period and more obvious oscillations. For the negative magnetic field $B=-2\pi\times \SI{0.5}{\kilo\hertz}$, the averaged energy $\langle H \rangle$ is negative and close to the bottom side of spectrum shown in Fig.~\ref{fig_AsymmetricSpectrum}(b) and (c). Actually, the initial states is basically dominant by the ground state, suppressing the oscillations.

Last, we show the energy spectra for different system sizes $N=4, 8, 12$ in Fig.~\ref{fig_Gapsofspectrum}.  For the system parameters $J_0=2\pi\times\SI{0.5}{\kilo\hertz}$, $\alpha=0.73$, there is always a non-vanishing energy gap in the high energy sector of the spectrum as increasing the number of ions. In the low energy sector of the spectrum, the level spacing decreases with system size increasing and the gap is expected to vanish in the thermodynamic limit. As a result, the energy gap at the high excited states leads to more persistent short-period oscillations while the persistent the oscillations at the ground state side are much suppressed and have long periods.

\end{document}